\documentclass
	[
	aps, pre
	, superscriptaddress 
	, preprint
	, amsmath, amssymb,amsfonts
	, floatfix
	, notitlepage
        ]
	 {revtex4-1}
\RequirePackage{graphicx,xcolor}
\graphicspath{{./figures/}} 
\RequirePackage{hyperref}

\newcommand{\of}[1]{\ensuremath{\left(#1\right)}}
\newcommand{\abs}[1]{\ensuremath{\left\vert#1\right\vert}}
\newcommand{\absStd}[1]{\ensuremath{\vert#1\vert}}
\newcommand{\fd}[1]{\ensuremath{\left[#1\right]}}
\newcommand{\set}[1]{\ensuremath{\left\lbrace#1\right\rbrace}}
\newcommand{\lsup}[1]{\ensuremath{^{\left(#1\right)}}}
\newcommand{\sgn}{{\mathrm{sgn}}}
\newcommand{\dl}[1][\;]{\ensuremath{\textrm{#1d}}}
\newcommand{\ft}[1]{\ensuremath{\hat{#1}}}
\newcommand{\x}[1][]{\ensuremath{\mathbf{r}_{#1}}}
\newcommand{\len}{\ensuremath{\mathit{l}}} 
\newcommand{\ar}{\ensuremath{\mathcal{A}}}
\newcommand{\V}{\ensuremath{V}} 
\newcommand{\vol}{\ensuremath{\mathcal{V}}}

\newcommand{\vc}[1]{\ensuremath{\boldsymbol{#1}}}
\newcommand{\nv}{\vc{n}}
\newcommand{\uv}{\vc{e}}
\newcommand{\film}{\ensuremath{\parallel}}
\newcommand{\twoSph}{\ensuremath{\circ\circ}}
\newcommand{\dSpace}{\ensuremath{D}} 
\newcommand{\dSp}{\ensuremath{d}} 

\newcommand{\aLab}{\ensuremath{A}} 
\newcommand{\bLab}{\ensuremath{B}} 
\newcommand{\pO}{\ensuremath{I}}
\newcommand{\pT}{\ensuremath{II}}
\newcommand{\T}{\ensuremath{T}}
\newcommand{\tcb}[1][]{\ensuremath{\T_{c}}}
\newcommand{\ca}[1][]{\ensuremath{{c}_{\aLab#1}}}
\newcommand{\cacb}[1][]{\ensuremath{\ca[,c]}}
\newcommand{\mf}{\ensuremath{\omega}}
\newcommand{\cP}{\ensuremath{\mu}}
\newcommand{\dcP}{\ensuremath{\Delta\cP}}
\newcommand{\dcPcb}{\ensuremath{\dcP_{c}}} 
\newcommand{\force}{\ensuremath{f}}
\newcommand{\stress}{\ensuremath{\mathcal{T}}}

\newcommand{\Fe}[1][]{\ensuremath{\mathcal{F}_{#1}}}
\newcommand{\fe}[1][]{\ensuremath{\mathfrak{f}_{#1}}}

\newcommand{\eosSig}{\ensuremath{E}}

\newcommand{\pot}[1][]{\ensuremath{U_{#1}}}

\newcommand{\Op}[1][]{\ensuremath{\phi_{#1}}}
\newcommand{\dOp}[1][]{\ensuremath{\partial_{#1}\Op}} 

\newcommand{\OpLf}[1][]{\ensuremath{\Phi_{#1}}}
\newcommand{\dOpLf}[1][]{\ensuremath{\partial_{#1}\OpLf}}

\newcommand{\OpBt}[1][]{\ensuremath{\mathcal{B}_{t#1}}}

\newcommand{\Ham}[1][]{\ensuremath{\mathcal{H_{#1}}}}
\newcommand{\ham}[1][]{\ensuremath{\mathtt{h}}}
\newcommand{\hb}[1][]{\ensuremath{h_{b#1}}}
\newcommand{\hs}[1][]{\ensuremath{h_{s#1}}}

\newcommand{\ccf}[1][]{\ensuremath{\force_{C#1}}}
\newcommand{\sff}[1][(\dSpace)]{\ensuremath{\vartheta^{#1}}}
\newcommand{\sfccfwght}[1][]{\ensuremath{\zeta_{\sff[]#1}}}

\newcommand{\CasAmpl}{\ensuremath{\Delta}}
\newcommand{\xit}[1][t]{\ensuremath{\xi_{#1}}}
\newcommand{\xih}[1][]{\ensuremath{\xi_{h#1}}}
\newcommand{\sfxi}[1][(\dSpace)]{\ensuremath{I^{#1}_{\pm}}}
\newcommand{\Sig}{\ensuremath{\mathsf{\Sigma}}}

\newcommand{\y}{\ensuremath{\mathcal{Y}}} 
\newcommand{\z}{\ensuremath{\mathsf{Y}}} 

\newcommand{\Fctl}[1][]{\ensuremath{\mathrm{F}_{#1}}}

\newcommand{\G}[1][]{\ensuremath{\mathcal{G}}}

\newcommand{\kb}{\ensuremath{k_B}}

\newcommand{\eref}[2][]{{Eq#1.~\eqref{#2}}}
\newcommand{\erefN}[4][]{[#2Eq#1.~\eqref{#3}#4]}
\newcommand{\fref}[2][]{{Fig#1.~\ref{#2}}}
\newcommand{\sref}[2][]{{Sec#1.~\ref{#2}}}
\newcommand{\rcite}[2][]{{Ref#1.~\cite{#2}}}

\usepackage{fancyhdr}
\begin{document}
\title 
   {Critical Casimir interactions around the consolute point of a binary solvent} 
\date{\today}
\author{T. F. Mohry}
\email{mohry@is.mpg.de}
\affiliation{Max-Planck-Institut f{\"u}r Intelligente Systeme,
  Heisenbergstra{\ss}e 3, 70569 Stuttgart, Germany}
\affiliation{IV. Institut f{\"u}r Theoretische Physik, Universit{\"a}t Stuttgart, 
  Pfaffenwaldring 57, 70569 Stuttgart, Germany}
\author{S. Kondrat}
\email{s.kondrat@fz-juelich.de}
\affiliation{Department of Chemistry, Imperial College London, SW7 2AZ, UK}
\affiliation{IBG-1: Biotechnology, Forschungszentrum J\"ulich, 52425 J\"ulich, Germany}

\author{A. Macio{\l}ek}
\email{maciolek@is.mpg.de}
\affiliation{Max-Planck-Institut f{\"u}r Intelligente Systeme,
  Heisenbergstra{\ss}e 3, 70569 Stuttgart, Germany}
\affiliation{IV. Institut f{\"u}r Theoretische Physik, Universit{\"a}t Stuttgart, 
  Pfaffenwaldring 57, 70569 Stuttgart, Germany}
\affiliation{Institute of Physical Chemistry, Polish Academy of Sciences,
  Kasprzaka 44/52, PL-01-224 Warsaw, Poland}
\author{S. Dietrich}
\email{dietrich@is.mpg.de}
\affiliation{Max-Planck-Institut f{\"u}r Intelligente Systeme,
  Heisenbergstra{\ss}e 3, 70569 Stuttgart, Germany}
\affiliation{IV. Institut f{\"u}r Theoretische Physik, Universit{\"a}t Stuttgart, 
  Pfaffenwaldring 57, 70569 Stuttgart, Germany}
\begin{abstract}
Spatial confinement of a near-critical medium changes its fluctuation spectrum 
and modifies the corresponding order parameter distribution. These effects result 
in effective, so-called critical Casimir forces (CCFs) acting on the confining 
surfaces. These forces are attractive for like boundary conditions of the order 
parameter at the opposing surfaces of the confinement. For colloidal particles 
dissolved in a binary liquid mixture acting as a solvent close to its critical 
point of demixing, one thus expects the emergence of phase segregation into 
equilibrium colloidal liquid and gas phases. We analyze how such phenomena occur 
asymmetrically in the whole thermodynamic neighborhood of the consolute point of 
the binary solvent. By applying field-theoretical methods within mean-field   
approximation and the semi-empirical de Gennes-Fisher functional, we study the
CCFs acting between planar parallel walls as well as between two spherical colloids 
and their dependence on temperature and  on the  composition of the  near-critical binary 
mixture. 
We find that for 
compositions slightly poor in the molecules preferentially adsorbed at the surfaces, 
the CCFs are significantly stronger than at the critical composition, thus leading 
to pronounced colloidal segregation. 
The segregation phase diagram of the colloid solution following from the calculated 
effective pair potential between the colloids agrees surprisingly well with experiments 
and simulations.
\end{abstract}
%
%
\maketitle
\section{Introduction \label{sec:intro}}
Finite-size contributions to the free energy of a spatially confined fluid 
give rise to an excess pressure, \textit{viz.}, an effective force per unit 
area acting on the confining surfaces. This so-called solvation force depends 
on the geometry of the confinement, the surface separation, the fluid-fluid 
interactions, the substrate potentials exhibited by the surfaces, and on the 
thermodynamic state of the fluid~\cite{Evans:1990}. 
The solvation force acquires a universal, long-ranged contribution upon approaching
the bulk critical point of the fluid, as first pointed out 
by Fisher and de Gennes~\cite{Fisher-et:1978}. 
This is due to \emph{critical order parameter fluctuations} which led to the 
notion of `critical Casimir forces', in analogy with the quantum-mechanical 
Casimir forces which are due to quantum fluctuations of confined electromagnetic 
fields~\cite{Casimir:1948}. 

The important role of critical Casimir forces (CCFs) for colloidal suspensions has 
implicitly been first recognized while studying experimentally aggregation phenomena 
in binary near-critical solvents~\cite{Beysens-et:1985}.
Numerous other experimental studies followed aiming to clarify important aspects of 
the observed phenomenon, such as its reversibility and the location of its occurrence 
in the temperature - composition phase diagram of the solvent (see, for example, 
\rcite[s]{Beysens-et:1999, *Gallagher-et:1992b, *Narayanan-et:1993, *Kurnaz-et:1997,%
Bonn-et:2009,*Gambassi-et:2010Ibid,*Bonn-et:2010Ibid,Veen-et:2012} and references therein). 
Measurements were  performed mostly in the homogeneous phase of the liquid mixture. They 
demonstrate that the temperature - composition $(T,c)$ region within which colloidal 
aggregation occurs is not symmetric about the critical composition $c_c$ of the
solvent mixture. Strong aggregation occurs on that side of the  critical composition which is 
rich in the component disfavored by the colloids. More recently, reversible fluid-fluid 
and fluid-solid phase transitions of colloids dissolved in the homogeneous phase of a 
binary liquid mixture have been observed \cite{Guo-et:2008,Nguyen-et:2013,Dang-et:2013}. 
These experiments also  show that the occurrence of such  phase transitions  is related 
to the affinity of the colloidal surfaces for one of the two solvent components as 
described above.

Various mechanisms for attraction between the colloids, which can lead to these 
phenomena, have been suggested. The role of dispersion interactions, which are 
effectively modified in the presence of an adsorption layer around the colloidal 
particles, has been discussed in Ref.~\cite{Law-et:1998}. 
A  ``bridging'' transition, which occurs when  the wetting films surrounding each 
colloid merge to form a liquid bridge \cite{Archer-et:2005}, provides a likely 
mechanism sufficiently off the critical composition of the solvent.
However, in the close vicinity of the bulk critical point of the solvent,
in line with the prediction by Fisher and de Gennes~\cite{Fisher-et:1978},
attraction induced by critical fluctuations should dominate. 
 
In the original argument  by Fisher and de Gennes,  the scaling analysis
for off-critical composition of the solvent  has not been carried out. Due to 
the lack of  explicit results for  the composition dependence of CCFs, for a 
long time it has not been possible to quantitatively relate the aggregation 
curves to CCFs. Rather, it was expected that CCFs play a negligible role 
for off-critical compositions because  away from  $c_c$  the bulk correlation 
length, which determines the range of CCFs, shrinks rapidly.  However, to a 
certain extent the properties of an aggregation region  {\it can}  be captured 
by assuming the attraction mechanism to be entirely due to CCFs. This has been 
shown in a recent theoretical study which employs an effective one-component 
description of the colloidal suspensions~\cite{Mohry-et:2012a,*Mohry-et:2012bIbid}. 
Such an approach is based on the assumption of additivity of CCFs 
and requires the knowledge  of the critical Casimir  pair potential  in the 
whole neighborhood of the critical point of the binary solvent, i.e., as a function of
both temperature and solvent composition close to $(T_c,c_c)$. 
In Ref.~\cite{Mohry-et:2012a,*Mohry-et:2012bIbid}, it was assumed that colloids are 
spheres all strongly preferring the same component of the binary mixture such that 
they impose symmetry breaking ($(+,+)$) boundary conditions \cite{Diehl:1997} 
on the order parameter of the solvent. 
Further, the pair potential between two spheres  has been 
expressed in terms of the scaling function of the CCFs  between two parallel plates 
by  using  the Derjaguin approximation \cite{Derjaguin:1934}.  
The dependence of the CCFs on the solvent composition translates into the dependence 
on the bulk ordering field $h_b$ conjugate to the order parameter 
(see Eq.~(A4) in the first part of Ref.~\cite{Mohry-et:2012a}).
For the parallel-plate (or film)  geometry in spatial dimension $\dSpace{}=3$, 
the latter has been approximated by  the  functional form  obtained within   
mean-field theory (MFT, $\dSpace{}=4$) by using the field-theoretical 
approach within the framework of Landau-Ginzburg theory.
The scaling functions of the CCFs resulting from these approximations have not 
yet been reported in the literature. We present them here for a wide range of 
parameters. In order to assess the quality of the approximations adopted in 
Ref.~\cite{Mohry-et:2012a,Mohry-et:2012bIbid} we calculate the scaling functions 
of the CCFs by using alternative theoretical approaches and compare the 
corresponding results.

In this spirit, one can estimate how well the  mean-field functional form, which 
is exact in $\dSpace{}=4$ (up to logarithmic corrections),  approximates the  
dependence on \hb{} of CCFs for films in $\dSpace{}=3$ by comparing it with the form 
obtained from  the local-functional approach~\cite{Fisher-et:1990a, *Fisher-et:1990bIbid} in 
$\dSpace{}=3$. We use the semi-empirical free energy functional developed by Fisher and 
Upton~\cite{Fisher-et:1990a, *Fisher-et:1990bIbid} in order to extend the original de 
Gennes-Fisher critical-point ansatz~\cite{Fisher-et:1978}.
Upon construction, this functional  fulfills the necessary analytic properties as 
function of $T$ and a proper scaling behavior for arbitrary \dSpace{}. The extended 
de Gennes-Fisher functional   provides results for CCFs in films with $(+,+)$ boundary 
conditions at $\hb=0$, which are in a good agreement with results from Monte Carlo 
simulations~\cite{Borjan-et:2008}.
A similar local-functional approach proposed by Okamoto and Onuki~\cite{Okamoto-et:2012} 
uses a renormalized Helmholtz free energy instead of the Helmholtz free energy of the 
linear parametric model used in Ref.~\cite{Borjan-et:2008}. Such a version does not 
seem to produce better results for the Casimir amplitudes~\cite{Okamoto-et:2012}. This 
`renormalized' local-functional theory has been recently applied to study the 
bridging transition between two spherical particles~\cite{Okamoto-et:2013}. 
Some results for the CCFs with strongly adsorbing walls and $h_b \ne 0$ obtained 
within mean-field theory and  within density functional theory in $\dSpace{}=3$
have been presented in Refs.~\cite{Schlesener-et:2003} 
and \cite{Buzzaccaro-et:2010, *Piazza-et:2011}, respectively.
These results are consistent with the present ones.  
 
We also explore the validity of the Derjaguin approximation for the mean-field scaling 
functions of the CCFs, focusing on their dependence on the bulk ordering field.  
For that purpose, we have performed bona fide mean-field calculations for spherical 
particles, the results of which can be viewed as exact for hypercylinders in $\dSpace=4$
or approximate for two spherical particles in $\dSpace=3$. 

This detailed knowledge of the CCFs as function of $T$ and $h_b$ is applied 
in order to analyze  recently published experimental data for the pair potential  and the segregation phase diagram \cite{Dang-et:2013}  
of poly-n-isopropyl-acrylamide microgel (PNIPAM) colloidal particles immersed in a 
near-critical 3-methyl-pyridine (3MP)/heavy water mixture.  

Our paper is organized such that 
in Sec.~\ref{sec:theory} we discuss the theoretical background.
In \sref{ssc:ccf_films}, results for CCFs for films are presented. These results 
as obtained from the field-theoretical approach within mean-field approximation 
are compared with those stemming from the local functional approach. 
We discuss how the dependence of the CCFs on the bulk ordering field \hb{} changes 
with  the spatial dimension \dSpace{}.
Section~\ref{ssc:ccf_spheres} is devoted to the CCF between spherical particles, 
where we also probe the reliability of the Derjaguin approximation. 
In Sec.~\ref{sec:exp} our theoretical results are confronted with the corresponding 
experimental findings and simulations.
We provide a summary in Sec.~\ref{sec:summary}.

\section{Theoretical background \label{sec:theory}}

For the demixing phase transition of 
a binary liquid mixture, the order parameter \Op{} is proportional to the 
deviation of the concentration $c = {\varrho}_{\aLab} - {\varrho}_{\bLab}$ from its 
value $c_c$ at the critical point, i.e., $\Op \sim c - c_{c}$; 
here ${\varrho}_{\alpha}$, $\alpha\in\set{\aLab,\bLab}$,  are the number densities of the 
particles of species \aLab{} and \bLab{}, respectively. The 
{\it{}b}ulk ordering field, conjugate to this order parameter, is  
proportional to the deviation of the difference $\dcP=\cP_{\aLab}-\cP_{\bLab}$ 
of the chemical potentials $\cP_{\alpha}$, $\alpha\in\set{\aLab,\bLab}$, of the 
two species from its critical value, i.e., $\hb \sim \dcP-\dcPcb$. 
We note, that the actual scaling fields for real fluids are 
linear combinations of \hb{} and the reduced temperature 
$t=\of{\tcb-\T}/\tcb$ [$t=\of{\T-\tcb}/\tcb$] 
for a lower [upper] critical point. 

Close to the bulk critical point, the bulk correlation length attains the scaling form
\begin{align}
\label{eq:td_crit_xiscaling}
  \xi\of{t,\hb}=\xit\sfxi\of{\abs{\Sig}=\xit/\xih}, 
\end{align}
where the \emph{universal} bulk scaling function \sfxi{}  satisfies  
$\sfxi\of{\abs{\Sig} \to 0} = 1$ and 
$\sfxi\of{\abs{\Sig} \to \infty} = \abs{\Sig}^{-1}$. 
The functional form of $\sfxi\of{\abs{\Sig}}$ depends on the 
sign \of{\pm} of $t$, but not on the sign of the 
bulk scaling variable \Sig. It is suitable to define the latter  
as $\sgn\of{\Sig}=\sgn\of{t\hb}$. 
The bulk correlation length for $\hb=0$ is 
\begin{subequations}
\label{eq:crit_corrlength}
    \begin{equation}
	\xit = \xit[{\pm}]\lsup{0} \abs{t}^{-\nu}
	\label{eq:crit_corrlength_xit}
    \end{equation}
    and  
    \begin{equation}
	\xih = \xih\lsup{0} \abs{\hb}^{-\nu/\of{\beta\delta}}
	\label{eq:crit_corrlength_xih}
    \end{equation}
\end{subequations}
is the bulk correlation length along the critical isotherm. 
Here $\nu$, $\beta$, 
and $\delta=\of{\dSpace\nu/\beta}-1$ are standard bulk critical exponents. 
For the Ising bulk  universality  class considered here,  $\nu=0.63$ and 
$\beta=0.33$ in spatial dimension $\dSpace=3$ and $\nu=\beta=1/2$ in 
$\dSpace\ge4$ ~\cite{Fisher:1967,Pelissetto-et:2002}. 
There are three non-universal amplitudes, $\xi_{\pm}\lsup{0}$ and 
$\xih\lsup{0}$, but the ratio  $U_{\xi}=\xi_{+}\lsup{0}/\xi_{-}\lsup{0}$ 
forms a universal number~\cite{Pelissetto-et:2002,NoteAmplRatios},
$U_{\xi}\of{\dSpace=3}\simeq 1.9$ and $U_{\xi}\of{\dSpace=4}=\sqrt{2}$. 
The values of $\xi_{\pm}\lsup{0}$ and $\xih\lsup{0}$ depend on the definition 
of $\xi$ which we take to be the true bulk correlation length governing the 
exponential decay of the two-point correlation function of the bulk order parameter. 
 
\subsection{Film geometry}
\label{sec:def:film}

For two parallel planar walls a distance $L$ apart 
the critical Casimir force is defined as~\cite{Brankov-et:2000,Gambassi:2009}
\begin{align}
\label{eq:ccf_def}
	\ccf\lsup{\film}=
		 -\frac{\partial \Fe[sgl]\lsup{ex}}{\partial L}
		 =-\frac{\partial \of{\Fe[sgl] -V \fe[b,sgl]}}{\partial L},
\end{align}
where \fe[b,sgl] is the singular part of the {\it{}b}ulk free energy density, 
$\Fe[sgl]\lsup{ex}$ is the singular part of the {\it{}ex}cess over the bulk 
free energy of  the film, and $V=\ar{}L$ where \ar{} is the macroscopically 
large surface area of one wall.

Finite-size scaling~\cite{Barber:1983} predicts that~\cite{Fisher-et:1978} 
\begin{align}
  \label{eq:ccf_film_scaling}
  \frac{\ccf\lsup{\film}}{\ar}= \frac{\kb\T}{L^{\dSpace}}\tilde\vartheta_{\film}^{(D)}
      (\y=\sgn\of{t}L/\xit,\Lambda=\sgn\of{\hb}L/\xih),
\end{align}
where \kb{} 
is the Boltzmann constant and  $\tilde\vartheta_{||}^{(D)}(\y,\Lambda)$ is a
\emph{universal} scaling function. Its functional form depends on the 
bulk universality class \emph{and} on the \emph{surface} universality 
classes of the confining walls. Here we focus on walls with the 
same adsorption preferences 
(expressed in terms of surface fields conjugate to the order parameter at the surfaces)
in the so-called strong adsorption limit in which 
$\Op\of{\x}\to\infty$ for the spatial coordinate $\x$ approaching 
the walls. Note that $\tilde\vartheta^{(D)}$ depends on the sign of $\hb$ because the 
surface fields at the confining walls break the bulk symmetry $\hb \to -\hb$.
Depending on the particular thermodynamic path under consideration, 
other representations of the scaling function of the critical Casimir force 
might be more convenient. For example, the scaling function 
$\hat\vartheta_{||}^{(D)}\left(\y=\sgn\of{t}L/\xit,\Sig=\sgn(th_b)\xi_t/\xi_h\right)$ 
lends itself  to describe the dependence of the CCFs on \hb{} at fixed temperature.
We will discuss the following representations 
\begin{equation}
\label{eq:ccf_scaling_representations}
\begin{split}%
    & \tilde\vartheta_{\film}^{(\dSpace)}(\y,\Lambda)
    = \hat\vartheta_{\film}^{(\dSpace)}
	(\y,\Sig=\frac{\Lambda}{\y} = \sgn(th_b)\frac{\xit}{\xih}) =
    \\
    & \bar\vartheta_{\film}^{(\dSpace)}(\Lambda,\Sig)
    =     \vartheta_{\film}^{(\dSpace)}
	  (\z=\frac{\y}{\sfxi\of{\abs{\Sig}}} = \sgn(t)\frac{L}{\xi},\Sig).
\end{split}%
\end{equation}

\subsection{Colloidal particles}
\label{sec:def:colloids}

We consider two spherical colloids, or more generally two hypercylinders, in spatial 
dimension \dSpace{}. A hypercylinder $H_{\dSpace,\dSp}$ has \dSp{} finite 
semiaxes of equal length $R$ and is translationally invariant in the remaining 
\of{\dSpace-\dSp} dimensions. 
Here, the two hypercylinders are assumed to be geometrically 
identical and aligned parallel to each other.  
We denote this geometry by \twoSph. 
For two hypercylinders at closest surface-to-surface distance 
$L$, the CCF $\ccf\lsup{\twoSph}$ is defined by the right hand side of 
\eref{eq:ccf_def} with  \Fe[sgl] as the singular contribution to the free 
energy of the binary solvent in the macroscopically large volume $V$ 
with two suspended colloids. 

The scaling function of the critical Casimir force between two hypercylinders 
$H_{\dSpace,\dSp}$, per ``length'' \len{} of the \of{\dSpace-\dSp}-dimensional 
hyperaxis, can be written as  \cite{Schlesener-et:2003, Kondrat-et:2009}
\begin{align}
\label{eq:ccf_spsp_scaling}
  \frac{\ccf\lsup{\twoSph}}{\len}=
    \frac{\kb\T} {L L^{\dSpace-\dSp}}
	\frac{1} {\Delta^{\of{\dSp-1}/2}}\sff[\of{\dSpace,\dSp}]_{\twoSph} 
	\of{\y=\sgn\of{t}\frac{L}{\xit},\Delta=\frac{L}{R},\Lambda=\sgn\of{\hb}\frac{L}{\xih}}.
\end{align} 

Within the Derjaguin approximation \cite{Derjaguin:1934} 
the total force between two spherical objects, $H_{3,3}$ or 
$H_{4,3}$, is taken  to be 
$ {\ccf\lsup{\twoSph}}/{\len}\simeq\int\mathrm{d}S\;\tilde{f}_C\lsup{\film}
  = 2\pi \int_0^R \mathrm{d}\rho\;\rho \tilde{f}_C\lsup{\film}\of{L\of{\rho}}$, 
where $\tilde{f}_C\lsup{\film}$ is the force per area  and 
$L\of{\rho}=L+2R\of{1-\sqrt{1-\of{\rho/R}^2}}$, leading to the scaling function
\erefN[s]{compare with }{eq:ccf_film_scaling}{ and \eqref{eq:ccf_spsp_scaling}} 
\begin{equation}
\label{eq:ccf_derj}
    \sff[\of{\dSpace,\dSp}]_{\twoSph,Derj}\of{\y,\Delta,\Lambda} 
    = \pi \int\limits_{1}^{1+2\Delta^{-1}}\mathrm{d}x \;
	x^{-\dSpace}\fd{1-\frac{\Delta}{2}\of{x-1}}
	\tilde\vartheta_{\film}^{(\dSpace)}\of{x\y,x\Lambda^{}}, \; \; (D,d) =
        (3, 3) \mbox{ and } (4, 3).
\end{equation}
 Note, that for $(D,d)=(4,4)$ in the expression for $\sff[(4,4)]_{\twoSph,Derj}$ 
 there is an additional factor of  $2\sqrt{(x-1)(1+\Delta(x-1)/4)}$ multiplying  the
integrand in \eref{eq:ccf_derj}.
Commonly \cite{Hanke-et:1998,Schlesener-et:2003,Kondrat-et:2009,
Troendle-et:2009,*Troendle-et:2010,*LabbeLaurent-et:2014,Hasenbusch:2013}, 
in this context [i.e., \eref{eq:ccf_derj}]  $\Delta$ is set to zero.                                                                       
Thus, within the Derjaguin approximation, 
$\ccf\lsup{\twoSph}\sim \Delta^{-\of{\dSp-1}/2}$  [\eref{eq:ccf_spsp_scaling}].  
We adopt this approximation except for, c.f., \fref{fig:FvsT_H0}(b), where we  shall 
discuss the full dependence on $\Delta$ given  by  \eref{eq:ccf_derj}.

\subsection{Landau theory} 
\label{sec:mft}
In the spirit of an expansion in terms of $\epsilon=4-\dSpace$, 
for the lowest order contribution we   use  the mean-field 
Landau-Ginzburg-Wilson theory in order to study the 
universal CCF in the film geometry (Sec.~\ref{ssc:ccf_films}) and between two 
colloidal particles (Sec.~\ref{ssc:ccf_spheres}). 
The Landau-Ginzburg-Wilson Hamiltonian, in units of $\kb\T$, 
is given by ~\cite{Binder:1983,Diehl:1986,Diehl:1997} 
\begin{equation}
\label{eq:ham_lg}
\Ham \fd{\Op\of{\x}} 
      =   \int_{\V} \set{
	  \frac{1}{2}\of{\nabla\Op}^2 
	+ \frac{\tau}{2}\Op^2 
	+ \frac{u}{4!}\Op^4 
	- \hb\Op
	} \dl^{\dSpace}r, 
\end{equation}
where \V{} is the volume of the confined critical medium, $\tau \propto t$ 
changes sign at the (mean-field) critical temperature \tcb, and the quartic 
term with the coupling constant $u>0$ stabilizes the Hamiltonian in the ordered 
phase, i.e., for $\tau<0$. Equation~\eqref{eq:ham_lg} must be supplemented by 
appropriate boundary conditions, which for the critical adsorption fixed point 
correspond to $\Op \to \pm \infty$.

Within mean-field theory, the bulk correlation lengths 
\erefN{}{eq:crit_corrlength}{} are~\cite{Schlesener-et:2003}
\begin{subequations}
\label{eq:xi_mft}
\begin{align}
\label{eq:xi_mft_xit}
    \xit\of{t>0} = \tau^{-1/2}, \quad  
    \xit\of{t<0} = \abs{2\tau}^{-1/2},
\end{align}
\begin{align}
\label{eq:xi_mft_xih}
    \xih\of{\hb\gtrless0} & = \abs{\sqrt{{9u}/{2}}\hb}^{-1/3}, 
\end{align}
\end{subequations}
and 
\begin{equation}
\label{eq:bulkcrit_xi_of_opb}
	\xi(t,\hb)= \set{ \fd{\xit\of{\abs{t}}}^{-2}\sgn\of{t} 
	  + (u/2) \Op[b]^2\of{t,\hb}}^{-1/2},
\end{equation}
where the bulk order parameter $\Op[b]\of{t,\hb}$ satisfies 
$\set{3\fd{\xit\of{\abs{t}}}^{-2}\sgn\of{t} +\frac{u}{2}\Op[b]^2}\sqrt{\frac{u}{2}}\Op[b]
=\of{\xih}^{-3}\sgn\of{\hb}$
so that $\frac{u}{2}\Op[b]^2$ can be expressed in terms of \xit{} and \xih{} 
and inserted into \eref{eq:bulkcrit_xi_of_opb}.
Within the present mean-field theory $\tau=\fd{\xit[+]\lsup{0}}^{-2}t$.

The minimum of \eref{eq:ham_lg} gives the {\it{}m}ean-{\it{}f}ield profile 
$\Op[mf]\of{\x;t,\hb}$.
With this the critical Casimir force is  
\begin{align}
    \label{eq:fc}
    \vc{\ccf} 	= {\kb \T} \int_{\ar} \stress (\Op) \cdot \nv \; d^{D-1} r
		= \len {\kb \T} \int_{\ar'} \stress (\Op) \cdot \nv \; d^{d-1} r
\end{align}
where \ar{} is an arbitrary \of{\dSpace-1}-dimensional surface enclosing a colloid or 
separating two planes, $\ar{}'$ is its \of{\dSp-1}-dimensional subset in the subspace 
in which the colloids have a finite extent, \nv{} is its unit outward normal, and  
\begin{equation}
\label{eq:stresstensor}
	\stress_{jk}\of{\Op}=
	    \frac{\delta \ham}{\delta \of{\dOp[k]}}\of{\dOp[j]}
	    -\delta_{jk}\ham
\end{equation}
is the stress tensor~\cite{Kondrat-et:2009}; here $\ham\of{\Op}$ 
is the integrand in \eref{eq:ham_lg}, and $\dOp[k]= \partial \Op / \partial x_k$. For 
the film geometry with chemically and geometrically uniform surfaces, the integration 
in \eref{eq:fc} amounts to the evaluation of $\stress$ at an arbitrary point between 
the two surfaces. For two spherical particles, the surface of integration is an arbitrary 
surface that encloses one of the particles. Accordingly, the force between the particles 
is $\vc{\ccf}=\ccf\uv$, where \uv{} is a unit vector along the line connecting 
their centers.
We have minimized the Hamiltonian 
\Ham{} numerically using the finite element method~\cite{f3dm}.  

Within mean-field theory, the scaling functions of the critical Casimir force can 
be determined only up to the prefactor $\sim 1/u$ (note that $u$ is dimensionless in 
$\dSpace=4$). In order to circumvent this uncertainty and to facilitate the comparison with 
experimental or other theoretical results, we shall normalize our mean-field results 
by the critical Casimir amplitude for the film geometry \erefN{see }{eq:ccf_film_scaling}{} 
$\Delta\lsup{4}_{\film}=\;\tilde\vartheta_{||}^{(D=4)}\of{\y=0,\Lambda=0}=-\of{6/u}{4}\fd{K\of{1/2}}^4 <0$
\cite{Krech:1997},
where $K$ is the complete elliptic integral of the first kind.
For the sphere-sphere geometry, one has 
\cite{Schlesener-et:2003,Kondrat-et:2009} 
$\sff[\of{\dSpace=4,\dSp=3}]_{\twoSph}\of{\y=0,\Delta=0,\Lambda=0}
 =\sff[\of{\dSpace=4,\dSp=3}]_{\twoSph,Derj}\of{\y=0,\Delta=0,\Lambda=0}
 =\frac{\pi}{3}\tilde\vartheta_{\film}^{(\dSpace=4)}\of {\y=0,\Lambda=0}
 =\frac{\pi}{3}      \vartheta_{\film}^{(\dSpace=4)}\of{\z=0,\Sig}$; 
note that $\Sig = \sgn(t\hb )\xit/\xih=const$ defines  implicitly various
thermodynamic paths $h_b(t)$ which, however,  all  pass the  critical point
 $(t=0, \hb=0)$, i.e., $\z=0$ [see Fig.~\ref{fig:sff4d_polar}]. Accordingly,   $\sff(\z=0,\Sig)$ does
 not depend on \Sig. 
Thus normalization by $\tilde\vartheta_{||}^{(D=4)}\of{0,0}$ eliminates the 
prefactor $\frac{1}{u}$. This holds also for nonzero values of \y{}, $\Delta$,  
and $\Lambda$ as well as beyond the Derjaguin approximation.

\subsection{Extended de Gennes-Fisher functional} 
\label{ssc:th_locfct}

For the film geometry, we consider the  \textit{ansatz} for the free energy functional 
proposed by Fisher and Upton \cite{Fisher-et:1990a,Fisher-et:1990bIbid} 
\begin{equation}
\label{eq:locfctal}
        \Fctl\fd{\OpLf\of{z}}=\ar{}\int_{-L/2}^{L/2}\left[\frac{\xi^2(\OpLf,t)}{2\chi(\OpLf,t)}(\dOpLf)^2
        +W\of{\OpLf,t,\hb}\right] \dl{z} 
          + \Fctl[s], 
\end{equation}
where $\dOpLf=\partial{\OpLf}/\partial{z}$. 
The equilibrium profile \OpLf[eq] is taken as the one which minimizes \Fctl{}. 
$\Fctl\fd{\OpLf[eq]}$ is the singular part of the free energy of the
near-critical medium confined in the film. 
Note that the order parameter \OpLf{} in 
\eref{eq:locfctal} is dimensionless, unlike \Op{} in the Landau model, in which it has 
the dimension $\of{\text{length}}^{1-\dSpace/2}$ \erefN{see }{eq:ham_lg}{}. 
The surface contribution $\Fctl[s]=-\hs[,1]\Op\of{z=-L/2}-\hs[,2]\Op\of{z=L/2}$ 
implements the boundary conditions. We consider walls adsorbing the same species
corresponding to surface fields $\hs[,1]=\hs[,2]>0$. 
$W\of{\OpLf;t,\hb}$ 
is the excess (over the bulk) free energy density (in units of $\kb T$), 
$\xi\of{\OpLf;t}$ and $\chi\of{\OpLf;t}$ are the bulk correlation length and 
the susceptibility of a homogeneous bulk system at \of{\OpLf,t}, 
respectively~\cite{Fisher-et:1990bIbid}.

Minimizing the functional given by Eq.~(\ref{eq:locfctal}) leads to an Euler-Lagrange equation, which can be 
formally integrated. One then proceed by taking  the scaling limit of this latter first integral and by  using 
the scaling forms
of the following bulk quantities: 
\begin{subequations}
	\label{eq:bulk_scaling}
	\begin{equation}
	\label{eq:W_scaling}
		W\of{\OpLf;t,\hb} 
		  =\abs{\OpLf}^{\delta+1}Y_{\pm}\of{\Psi,\Sig} \\
	\end{equation}
	{and}
	\begin{equation}
	\label{eq:Z_scaling}
		\xi^2/\of{2\chi} 
		  =\abs{\OpLf}^{\eta\nu/\beta}Z_{\pm}\of{\Psi},
	\end{equation}
\end{subequations}
where $\Psi=\OpLf/\OpLf[b]\of{-\abs{t},\hb=0}=\abs{t}^{-\beta}\OpLf/\OpBt$. The 
(dimensionless) non-universal amplitude \OpBt{} of the bulk order parameter \OpLf[b] can be 
expressed via universal amplitude ratios in terms of the non-universal amplitudes 
$\xit[+]\lsup{0}$ and $\xit[h]\lsup{0}$ of the bulk 
correlation length~\cite{Pelissetto-et:2002}; 
 the functions 
$\tilde{Y}_{\pm}\of{\Psi,\Sigma}=Y_{\pm}\of{\Psi,\Sigma}/Y_{+} \of{\infty,0}$ and 
$\tilde{Z}_{\pm}\of{\Psi}=Z_{\pm}\of{\Psi}/Z_{+}\of{\infty}$  are universal.
This procedure determines  (even without knowing the explicit
functional forms of $\tilde{Y}_{\pm}$ and $\tilde{Z}_{\pm}$) 
 a formal expression for the scaling function $\hat\vartheta_{||}^{(\dSpace)}$ of the CCF 
\cite{Borjan:1999,Borjan-et:2008}.
Here we take into account the additional dependence on the scaling variable 
$\Sig{}\neq 0$ and obtain for $\y>0$ \cite{Mohry:2013}
\begin{equation}
  \label{eq:ccf_scalfct_locfct}
  \hat\vartheta_{||}^{(\dSpace)}\of{\y>0,\Sig} = 
	-A_1 \abs{\y}^{2-\alpha}{\Psi_{m}}^{\of{1+\delta}}
	\tilde{Y}_{+}\of{\Psi_{m},\Sig},
\end{equation}
where $A_1=R_{\chi} Q_c/\of{\delta+1}$ is a universal number which is expressed in 
terms of the universal amplitude ratios~\cite{Pelissetto-et:2002,NoteAmplRatios} 
$R_{\chi}$, $Q_2$, and $Q_c$. $\Psi_{m}$ is defined through 
$\OpLf[m]=\OpLf (z =z_{m}) = \Psi_{m} \OpLf[b] (-\abs{t},\hb=0)$, which
for the present case $\hs[,1],\hs[,2]>0$
is the {\it{}m}inimal value of the order parameter profile across the film.

In order to calculate the critical Casimir force from \eref{eq:ccf_scalfct_locfct}  one 
has to evaluate the functions $Y_{\pm}$ and $Z_{\pm}$ in \eref{eq:bulk_scaling}. 
The analytical expressions of these functions can be obtained by using  
the so-called  linear parametric 
representation \cite{Schofield:1969,*Josephson:1969,*Fisher:1971,Pelissetto-et:2002,Borjan-et:2008}.
For given $\y$ and $\Sig$ the scaling function of the critical Casimir force is 
then computed numerically (for details see  Ref.~\cite{Mohry:2013}).

\section{Numerical results \label{sec:results}}

\subsection{Critical Casimir forces in films \label{ssc:ccf_films}}

\begin{figure}
        \includegraphics{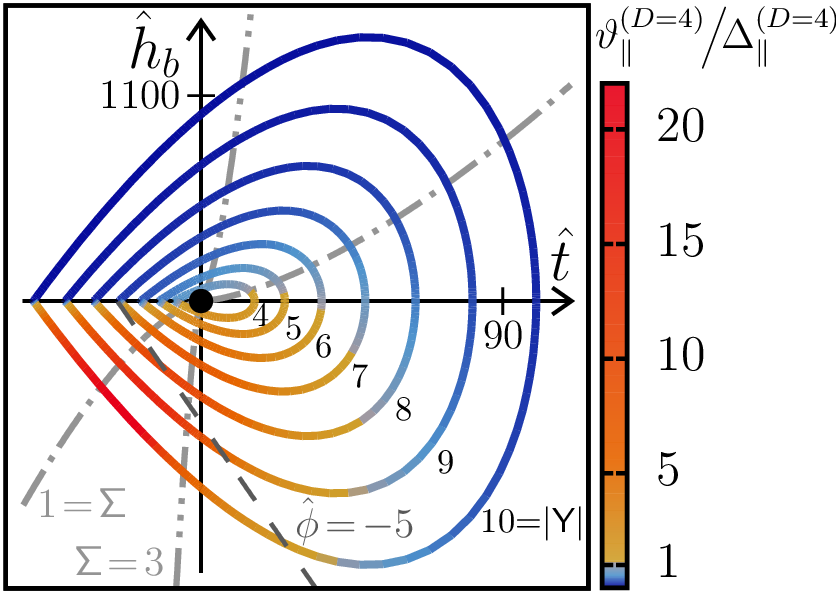}
        \caption{%
	Behavior of the normalized  mean-field Casimir scaling 
	function $\sff[\of{\dSpace=4}]_{\film} \of{\z=\sgn\of{t}L/\xi\of{t,\hb},\Sig=\sgn\of{t\hb}\xit/\xih}$
	 along lines of constant scaling variable  
	$\abs{\z}=4,5,\ldots, 10$  (from the inner to the outermost ring) 
	 in the thermodynamic  state space   of the  solvent spanned by 
	 $\hat t = \of{L/\xit[+]\lsup{0}}^{1/\nu}t$ and 
	$\hat{h}_b = \of{L/\xih\lsup{0}}^{\beta\delta/\nu}h_b$. 
	The color along the lines of constant \abs{\z} indicates the 
	absolute value \absStd{\sff[\of{\dSpace=4}]_{\film}}.  
	 The \emph{bulk} critical point of the solvent $(\hat t,\hat{h}_b)=(0,0)$ is 
         indicated by $\bullet$. The region  shown here lies {\it above}
	 the capillary transition critical point
	 \of{\z_{\film,c}=-11,\Sig_{\film,c}=1.3} \cite{Schlesener-et:2003},
	 where the \emph{film} coexistence line ends.
	 For $(+,+)$ boundary conditions, the  capillary condensation transition occurs 
	 for  $\hat t<0$ and $\hat{h}_b<0$.	 %
	The dash-dotted lines indicate the thermodynamic paths $\Sig=1$ and $=3$
	and the dashed line the path of constant order parameter 
        $\hat\Op=\frac{L}{\xit[+]\lsup{0}}\Op/\OpBt=-5$.
         Within mean-field theory
	$\nu=1/2$ and $\nu/\of{\beta\delta}=1/3$. 
	$\CasAmpl_{\film}^{\of{\dSpace=4}}=
	  \absStd{{\vartheta}^{\of{\dSpace=4}}_{\film}\of{\z=0,\Sig}}$, 
	which is independent of $\Sig$. 
	}
	\label{fig:sff4d_polar}
\end{figure}

Our mean-field results for  the behavior of the  Casimir scaling function
around the consolute point of the binary solvent are summarized in Fig.~\ref{fig:sff4d_polar}
in terms of  the    scaling function $\sff[\of{\dSpace=4}]_{\film}
  \of{\z=\sgn\of{t}L/\xi\of{t,\hb},\Sig=\sgn\of{t\hb}\xit/\xih}$. 
This particular  scaling form turns out to be 
particularly suitable in view of the Derjaguin 
approximation used below  for the sphere-sphere geometry because the 
dependence of the CCF on $L$, measured in units  of the true bulk correlation 
length  $\xi\of{t,\hb}$, enters $\sff[\of{\dSpace=4}]_{\film}$  only via $\z$.
The second scaling variable $\Sig$, which depends on the thermodynamic state 
of the solvent, varies smoothly from $\Sig = 0$ at the bulk coexistence 
curve to $\Sig = \pm\infty$ at the critical isotherm.

In Fig.~\ref{fig:sff4d_polar}, we have plotted several lines of constant
scaling variable $\abs{\z}=L/\xi\of{t,\hb}=4,5,\ldots,10$ in the 
thermodynamic space of the solvent  spanned by  $\hat t = \of{L/\xit[+]\lsup{0}}^{1/\nu}t$ and 
	$\hat{h}_b = \of{L/\xih\lsup{0}}^{\beta\delta/\nu}h_b$. 
The shape of the lines $\abs{\z}=\mathrm{const}$ is determined by the bulk 
correlation length $\xi(t,h_b)$. Therefore it
is symmetric about the  $\hat t$-axis. A break of slope occurs at the bulk coexistence line  
\of{\hat t<0,\hat{h}_b=0}  because  $\xi\of{t,\hb}$  depends on the bulk order parameter \Op[b] 
\erefN{see }{eq:bulkcrit_xi_of_opb}{} which varies there discontinously.
We use the color code to indicate  the strength 
\absStd{\sff[\of{\dSpace=4}]_{\film}} of the Casimir scaling function along these lines.
For \of{+,+} boundary conditions the critical Casimir force in a slab 
is attractive and accordingly $\sff[\of{\dSpace=4}]_{\film}<0$ for all values of 
$t$ and $\hb$. 

The main message conveyed by Fig.~\ref{fig:sff4d_polar}  is  the  
asymmetry  of  the critical Casimir force around the critical point
of the solvent with  the  maximum strength  occurring at $\hb<0$.
This asymmetry is due to the presence of  surface fields which break 
the  bulk symmetry $\hb\to-\hb$ of the system and shift the  phase  coexistence
line away from the bulk location $h_b=0$. In the film with $(+,+)$ boundary conditions 
the shifted, so-called capillary condensation transition,
occurs for  negative values of $h_b$ \cite{Evans:1990, Nakanishi-et:1983}. 
At capillary condensation, the solvation force (which within this context 
is a more appropriate notion than the notion of  the critical Casimir force) 
exhibits a jump from a large value for thermodynamic states corresponding to the \of{+} phase
to a vanishingly small value for those corresponding to the \of{-} phase.
Above  the two-dimensional plane spanned by $(\hat{t}, \hat{h}_b)$, 
the surface $\sff[\of{\dSpace=4}]_{\film}$ 
forms a trough which is the remnant of these jumps extending 
to the thermodynamic region above the capillary condensation 
critical point, even to temperatures higher than \tcb. This trough, reflecting 
the large strengths visible 
in Fig.~\ref{fig:sff4d_polar} for $\hat{h}_b<0$,  deepens upon approaching 
the capillary condensation point.

Along the particular thermodynamic path of zero bulk field (i.e., $\Sig=0$)
the minimum is located  above $T_c$ and has the value   
$\sff[\of{4}]_{\film}\of{\z_{min}=3.8,\Sig=0}
    = 1.4\times{} \sff[\of{4}]_{\film}\of{0,0}$.
Along the critical isotherm (i.e., $\abs{\Sig}=\infty$) one has
$\sff[\of{4}]_{\film}\of{\z_{min}=8.4,\Sig=-\infty}
    =  10\times{} \sff[\of{4}]_{\film}\of{0,0}$.
Interestingly, along all lines $\abs{\z}=L/\xi\of{t,\hb}=\mathrm{const}$  
the strength \absStd{\sff[\of{4}]_{\film}\of{\z=\mathrm{const},\Sig}} 
takes its minimal value at the bulk coexistence curve $\hb=0^+$.
For $\abs{\z}\gtrless6.3$ the maximal value of 
\absStd{\sff[\of{4}]_{\film}\of{\z=\mathrm{const},\Sig}}
is located at $\hb<0$ and $t\lessgtr0$.

It is useful to consider the variation of the scaling function of the CCF along  
the thermodynamic paths of fixed $\Sig$. As examples, such paths are shown 
for $\Sig=1$ and $\Sig=3$ in Fig.~\ref{fig:sff4d_polar} as dash-dotted  lines. 
Thermodynamic paths corresponding to $0 < \Sig \lesssim 1.3 $ cross the phase 
boundary of {\it{}c}oe{\it{}x}isting phases in the film at certain values 
$\z_{cx}\of{\Sig}$,  which lie outside the range of the plot in Fig.~\ref{fig:sff4d_polar}. 
Along the paths corresponding to $0<\Sig\lesssim 3$,  
$\sff[\of{4}]_{\film}\of{\z,\Sig=const<3}$ as function of \z{} has two minima. 
The  local minimum  occurs above $T_c$,  whereas  the global one occurs below  $T_c$.
For all other fixed values of \Sig, the scaling function  $\sff[\of{4}]_{\film}$,  
as function of \z{},  exhibits a single minimum;
for negative \Sig{}  it is located above $T_c$ (i.e., $\z{}>0$), whereas for 
$\Sig \gtrsim 3 $ below $T_c$ (i.e., $\z{}<0$).
Results for $\sff[\of{\dSpace=4}]_{\film}$ as function of 
$\z=\sgn\of{t}L/\xi\of{t,\hb}$ for constant values of 
$\Sig=\sgn\of{t\hb}\xit/\xih$ are shown in \rcite{Mohry:2013}.  

Thermodynamic paths of constant order parameter $\Op\neq 0$ are particularly experimentally 
relevant, because they correspond to a fixed off-critical composition of the solvent.  
As an example \fref{fig:sff4d_polar} shows the case 
$\hat{\phi} = \frac{L}{\xit[+]\lsup{0}}\phi/\OpBt=-5$ as
indicated by the dashed line. 
Within mean-field theory this path varies linearly with $t$.

\begin{figure}
        \includegraphics{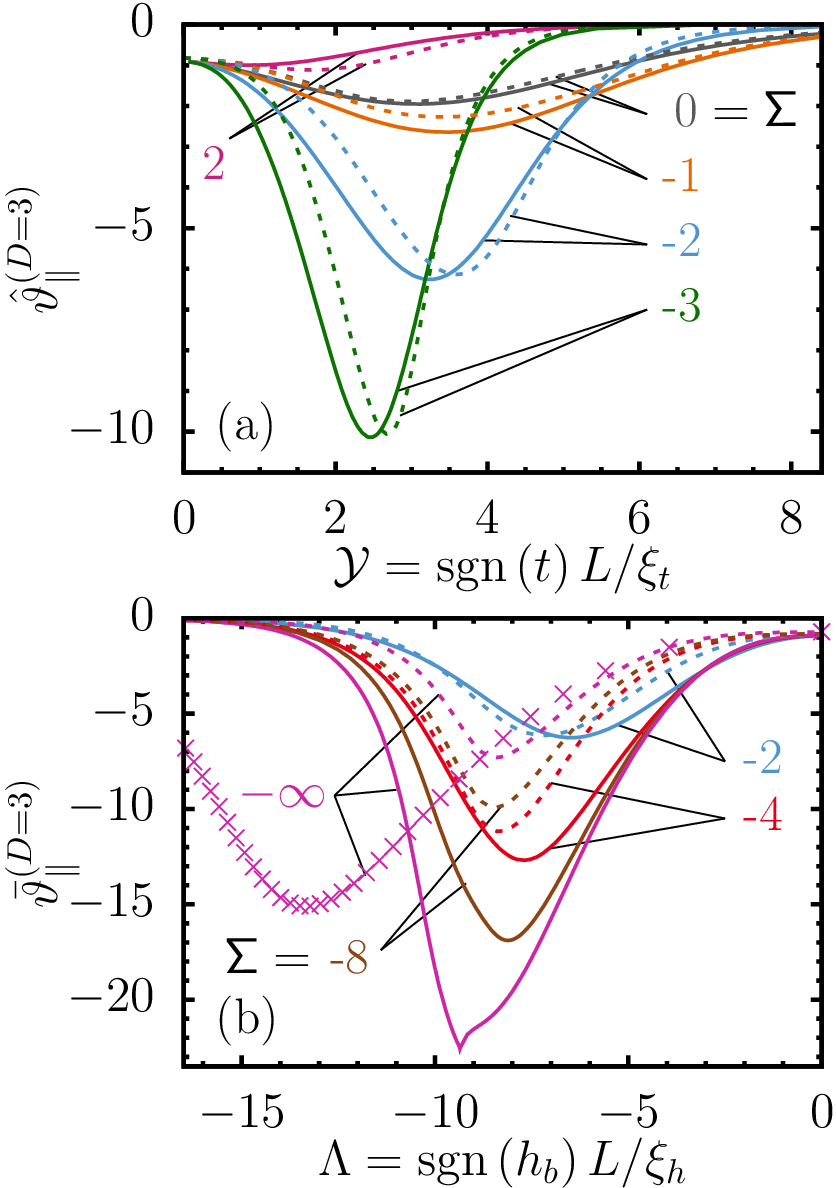}
        \caption{ 
	  Two representations of the scaling function of the critical Casimir force 
	  for the film geometry with $(+,+)$ boundary conditions 
	  \erefN{}{eq:ccf_scaling_representations}{}:
	   (a) $\hat\vartheta^{(\dSpace=3)}_{\film}\of{\y,\Sig}$  plotted \textit{versus}   $\y=\sgn\of{t}L/\xit$ and 
	   (b) $\bar\vartheta^{(\dSpace=3)}_{\film}\of{\Lambda,\Sig}$ 
	    plotted \textit{versus} $\Lambda=\sgn\of{\hb}L/\xih$
	   for several values of the scaling variable
	  $\Sig{} =\sgn\of{t\hb}\xit/\xih$.
	  The full lines are the results obtained from the local functional approach together 
	  with the linear parametric model \erefN{}{eq:ccf_scalfct_locfct}{},  
	  and the dashed lines correspond to the dimensional approximation 
	  \erefN{}{eq:sff_h_approx}{}.
	  In (b)  the  symbols are Monte Carlo data from \rcite{Vasilyev-et:2013} 
	  for $\Sig = -\infty$. We note that  thermodynamic states corresponding to
	  $\Sig = -\infty$ and $\Sig = \infty$  are the same; they form the  critical isotherm.}
        \label{fig:sff3d} 
\end{figure}
In \rcite{Mohry-et:2012a,*Mohry-et:2012bIbid},  the mean-field results described above
were used in order to approximate
the dependence of the CCFs on the bulk ordering field \hb{} in spatial dimension $\dSpace=3$:
\begin{equation}
\label{eq:sff_h_approx}
    \hat\vartheta^{(\dSpace)}_{\film}\of{\y,\Sig}\simeq
		\hat\vartheta^{(\dSpace)}_{\film}\of{\y,\Sig=0}
		\frac{\hat\vartheta^{(\dSpace'=4)}_{\film}\of{\y,\Sig}}
		     {\hat\vartheta^{(\dSpace'=4)}_{\film}\of{\y,\Sig=0}},
\end{equation}
where $\hat\vartheta^{(\dSpace=3)}_{\film}\of{\y,\Sig=0}$ is 
taken from Monte Carlo simulation data \cite{Vasilyev-et:2009}. 
This ``dimensional approximation'' is inspired by the observation that the trends 
and qualitative features of 
$\hat\vartheta^{(\dSpace)}_{\film}$ are the same for different values of \dSpace{} 
\cite{Vasilyev-et:2007,*Vasilyev-et:2009, Mohry-et:2010}.
The characteristics of this approximation are as follows:
(i) For $\dSpace\to\dSpace'=4$, i.e., for mean-field theory, the right hand side of 
\eref{eq:sff_h_approx} turns into the correct expression for the full range of 
all scaling variables. 
(ii) For $\hb\to 0$ (i.e., $\Sig\to0$) the right hand side of 
\eref{eq:sff_h_approx} reduces exactly to
$\hat\vartheta^{(\dSpace)}_{\film}\of{\y,\Sig=0}$ for all values $\dSpace$, $\dSpace'$, and $\y$.  
In this sense the approximation is concentrated on the dependence on \hb.
(iii) For $\dSpace'=4$ the approximation can be understood as the lowest order 
contribution in an $\epsilon=4-\dSpace$ expansion of 
$\sfccfwght[{}_{\film}]\lsup{\dSpace}=      
	{\hat\vartheta^{(\dSpace)}_{\film}\of{\y,\Sig}}/
	 {\hat\vartheta^{(\dSpace)}_{\film}\of{\y,\Sig=0}} $ which carries the 
whole dependence of the CCFs on $\Sig$. 
As a ratio, the mean-field expression for   
$\sfccfwght[{}_{\film}]\lsup{\dSpace'=4}$ does not suffer from the 
amplitude of $\hat\vartheta^{(\dSpace'=4)}_{\film}$ being undetermined.
In \eref{eq:sff_h_approx}, the scaling variables
$\y$ and $\Sig$ are taken to involve the critical bulk exponents in spatial dimension 
$\dSpace$ so that the approximation concerns only the shape of the scaling function.
The use of  bulk critical exponents in spatial dimension \dSpace{} for 
scaling variables which, however, are arguments of the scaling function in spatial
dimension $\dSpace' \neq \dSpace $, may lead to a deviation from the proper asymptotic 
behavior. 
However, this potential violation of the proper asymptotic behavior 
of the scaling function of the CCFs is expected to  
occur for large values of the arguments of the scaling function for which its 
value is exponentially small. Thus, the potential violation should not 
matter quantitatively in the range of the values of \y{} and \Sig{} for which the scaling function 
$\hat\vartheta^{(\dSpace)}_{\film}$ attains noticeable values. 
Here we compare this approximation  with the results obtained from the extended 
de Gennes-Fisher functional using the linear parametric model.

In \fref{fig:sff3d}(a) we plot   $\hat\vartheta_{||}^{(D=3)}(\y,\Sig)$ 
as a function of $\y > 0$ for several values of $\Sig$. 
For large values of \abs{\Sig} the relevant part of the corresponding 
thermodynamic path is close to the critical isotherm and accordingly the 
scaling variable $\Lambda=\Sig\y=\sgn\of{\hb}L/\xih$ is more appropriate
than the scaling variable \y. Therefore, in \fref{fig:sff3d}(b) we show 
$\bar\vartheta_{||}^{(D=3)}(\Lambda,\Sig)
=\hat\vartheta_{||}^{(D=3)}(\Lambda/\Sig,\Sig)$ 
as a function of $\Lambda$ for several fixed  values of $\Sig \le -2$.

As can be inferred from \fref{fig:sff3d} the dimensional approximation 
in \eref{eq:sff_h_approx} works well for weak bulk fields (such that $\abs{\Sig}<3$). 
Although the minima of the scaling functions are slightly 
shifted relative to each other, the depths 
of these minima   compare well  with the results of the
local functional approach. 
For all $\abs{\Sig}<\infty$, the value $\y=0$ corresponds to 
the bulk critical point and thus at $\y=0$ the curves $\hat\vartheta^{(\dSpace)}_{\film}$ 
attain the same value [see \fref{fig:sff3d}(a)].

For strong bulk fields, i.e., $\Sig<-4$ the dimensional approximation 
\erefN{}{eq:sff_h_approx}{} fails [see \fref{fig:sff3d}(b)]. For example,   
$|\bar\vartheta^{(\dSpace=3)}_{\film}|$ of the approximative curve 
becomes smaller for more negative values of \Sig{}, which is  
in contrast to the results of mean-field theory and of the local functional approach.  
This wrong trend of the results of the dimensional approximation is 
explained in detail in \rcite{Mohry:2013}.

We note that  the scaling functions $\sff[\of{\dSpace=3}]_{\film}$ of the critical Casimir 
force as obtained from the local functional exhibit the same qualitative 
features as the ones calculated within mean-field theory.
For example, the position $\z_{min}\of{\Sig}$ of the minimum as obtained from the 
present local functional theory changes from 
$\z_{min}\of{\Sig=0}=\y_{min}\of{\Sig=0}= 3.1$ 
at the thermodynamic path $\hb=0$ towards 
$\z_{min}\of{\Sig=-\infty}=-\Lambda_{min}\of{\Sig=-\infty}= 9.4$ 
at the critical isotherm. These values are similar to the ones 
obtained from mean-field theory.
The results  of the local functional approach  are peculiar with respect to 
the cusp-like minimum for curves close to the critical isotherm 
[for $\abs{\Sig}=\infty$, i.e., $t=0$, see \fref{fig:sff3d}(b)].
Such a  behavior  is also reported for the similar approach 
used in \rcite{Okamoto-et:2012}.
However,  there is no such  cusp  in the Monte Carlo data 
for $t=0$, i.e., $\abs{\Sig}=\infty$, \cite{Vasilyev-et:2013}
[see the symbols  in \fref{fig:sff3d}(b)].
As compared with the results of the local functional, the 
minimum of $\bar{\vartheta}^{\of{\dSpace=3}}_{\film}(\Lambda,\abs{\Sig}=\infty)$ 
obtained from Monte Carlo simulations is less deep and  is   
positioned at a more negative value of $\Lambda$. 
For $\Lambda>0$ [not shown in \fref{fig:sff3d}(b)], 
$\bar{\vartheta}^{\of{\dSpace=3}}_{\film}(\Lambda,\abs{\Sig}=\infty)$
as obtained from the local functional is less negative than the 
corresponding scaling function obtained from 
the Monte Carlo simulations.

We observe that  upon decreasing the spatial dimension \dSpace{} the ratio 
of  the strengths \abs{\hat\vartheta^{(D)}_{\film}} at its two extrema, the one located 
at the critical isotherm and the other located at  $h_b=0$,  increases, 
from $7$ in $\dSpace=4$ to $11.5$ (local functional) or
 $8$ (Monte Carlo simulations) in $\dSpace=3$, and  
to $15$ in $\dSpace=2$ \cite{Drzewinski-et:2000b}.

\subsection{Critical Casimir forces between spherical colloids} 
\label{ssc:ccf_spheres}

The CCF between two spherical colloids takes the form 
given by \eref{eq:ccf_spsp_scaling}; here we take $d=3$ and $D=4$.
In order to calculate $\sff[\of{4,3}]_{\twoSph}$, we use the stress tensor
 $\stress(\Op)$ \erefN[s]{see }{eq:fc}{ and \eref{eq:stresstensor}} 
with the mean-field profile $\Op(\x)$ which is determined by minimizing 
the Hamiltonian in \eref{eq:ham_lg} numerically
using GSL~\cite{gsl} and F3DM~\cite{f3dm} defined on  
three-dimensional meshes generated by TETGEN ~\cite{tetgen}.

CCFs between spherical colloids in zero bulk field have
been widely studied in the literature 
\cite{Hanke-et:1998,Schlesener-et:2003,Burkhardt-et:1995, Hasenbusch:2013}.
Whereas so far the mean-field theory of the Landau model has only been
considered for four-dimensional spheres $H_{D=4,d=4}$,
here we focus on three-dimensional spheres, i.e., on 
hypercylinders $H_{3,3}$ or $H_{D \ge 4,3}$.
We first consider the corresponding CCFs in zero bulk field \of{\Lambda=0}. 
We recall that we consider $(+,+)$ boundary conditions only.

\begin{figure}[!t]
    \includegraphics[width=0.47\textwidth]{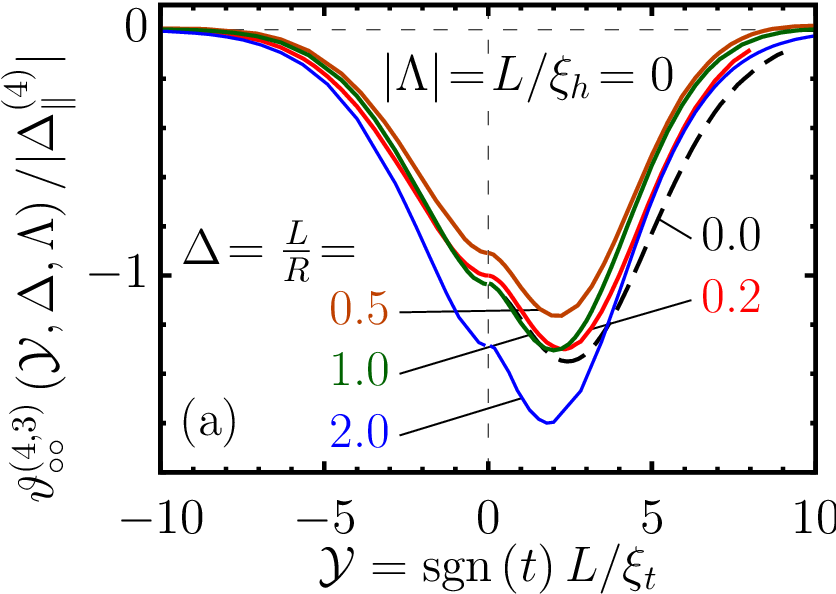}
    \includegraphics[width=0.47\textwidth]{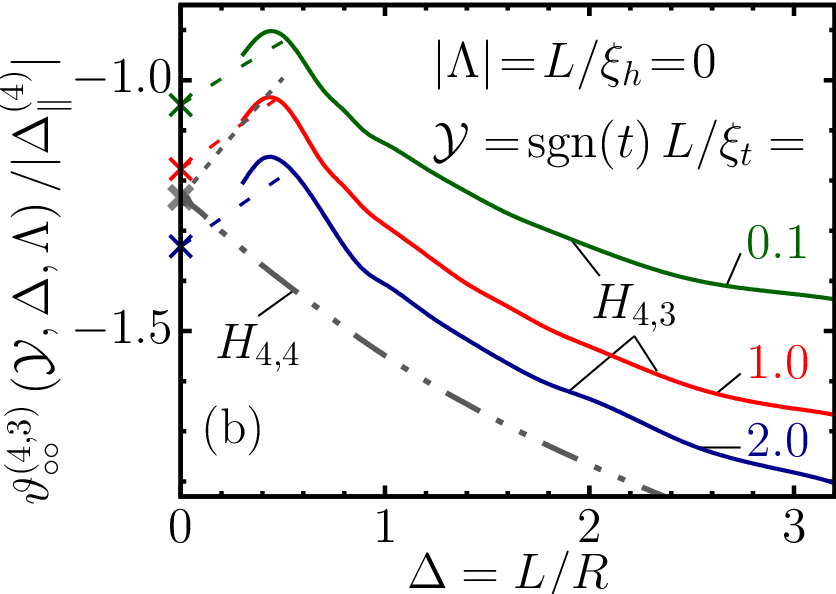}
    \caption{
	The critical Casimir force between two like colloids in zero bulk
	field ($\Lambda=0$) as obtained from the mean-field theory of the Landau model.
	(a)~The normalized  scaling function $\sff[\of{\dSpace=4,\dSp=3}]_{\twoSph}$
	\textit{versus} $\y = \sgn\of{t} L/\xit$ for five values of $\Delta = L/R$, where 
	$L$ is the surface-to-surface distance between two spheres $H_{4,3}$ of radius $R$.
	The curve $\Delta = 0$ corresponds to the Derjaguin approximation.
	(b)~The normalized scaling function $\sff[\of{4,3}]_{\twoSph}$ \textit{versus} $\Delta$
	for three values of $\y > 0$.
	The results of the Derjaguin approximation as given by \eref{eq:ccf_derj} 
	are shown as dashed lines ($\Delta\to0$) and by crosses ($\Delta=0$). 
	We recall the relation 
	$\sff[\of{\dSpace=4,\dSp=3}]_{\twoSph}\of{\y=0,\Delta=0,\Lambda=0}
	 =\frac{\pi}{3}\tilde\vartheta_{\film}^{(\dSpace=4)}\of{0,0}
	 =\frac{\pi}{3}\Delta_{\film}^{(4)}$.
	For comparison, the scaling function $\sff[\of{\dSpace=4,\dSp=4}]_{\twoSph}$ 
	at the critical point ($\y=\Lambda=0$) for spheres $H_{4,4}$~\cite{Schlesener-et:2003}
	is shown by the grey dash-dotted line. The  Derjaguin approximation for $H_{4,4}$
	(corresponding to the grey dotted line emerging  from
	the grey cross) displays, as a function of $\Delta$,  a trend opposite
        to the result 
	$\sff[\of{\dSpace=4,\dSp=4}]_{\twoSph}$ obtained from  the full calculation.
	}

   \label{fig:FvsT_H0}
\end{figure}

The scaling function
$\sff[\of{4,3}]_{\twoSph}\of{\y,\Delta=\mathrm{const},\Lambda = 0}$, as a
function of $\y$,
has a shape which is typical for like  boundary conditions [see
\fref{fig:FvsT_H0}(a)]. Interestingly, 
the magnitude of $\sff[\of{4,3}]_{\twoSph}$ depends \emph{non-monotonically} on
$\Delta$. This is shown explicitly in \fref{fig:FvsT_H0}(b),
where the scaling function is plotted \textit{versus} $\Delta$ for three 
values of $\y = \sgn(t)L / \xit > 0$. In \fref{fig:FvsT_H0}(b), $\sff[\of{4,3}]_{\twoSph}$
approaches the scaling function of the Derjaguin approximation from above
when $\Delta \to 0$, 
but decreases upon increasing $\Delta > \Delta_m$, where $\Delta_m \approx 1/2$ seems 
to be almost independent of $\y$ (in the range of $\y$ shown).
This non-monotonic behavior is unlike the case of four-dimensional spheres 
in $D=4$ dimension (i.e., $H_{4,4}$ hypercylinders), for which the scaling function 
approaches its value at $\Delta=0$ from below 
and exhibits no maxima (grey dash-dotted line in \fref{fig:FvsT_H0}(b) 
reproduced from Ref.~\cite{Schlesener-et:2003}). 
For the wall-sphere geometry, such a non-monotonic behavior of the scaling function 
of the CCF for $\Delta\to0$ has been found for a sphere $H_{3,3}$  using  Monte 
Carlo simulations \cite{Hasenbusch:2013}, 
but not for (hyper)cylinders $H_{4,d}$, $d\in\set{2,3}$, treated by mean-field theory 
\cite{Troendle-et:2009,*Troendle-et:2010,*LabbeLaurent-et:2014}.

The behavior of $\sff[\of{4,3}]_{\twoSph}$ for large $\Delta \gg 1$ is not
quite clear due to technical difficulties associated with large mesh sizes 
and the increasing numerical inaccuracy; moreover, in this limit, the force 
attains very small values.  

\begin{figure}[!t]
    \includegraphics[width=0.47\textwidth]{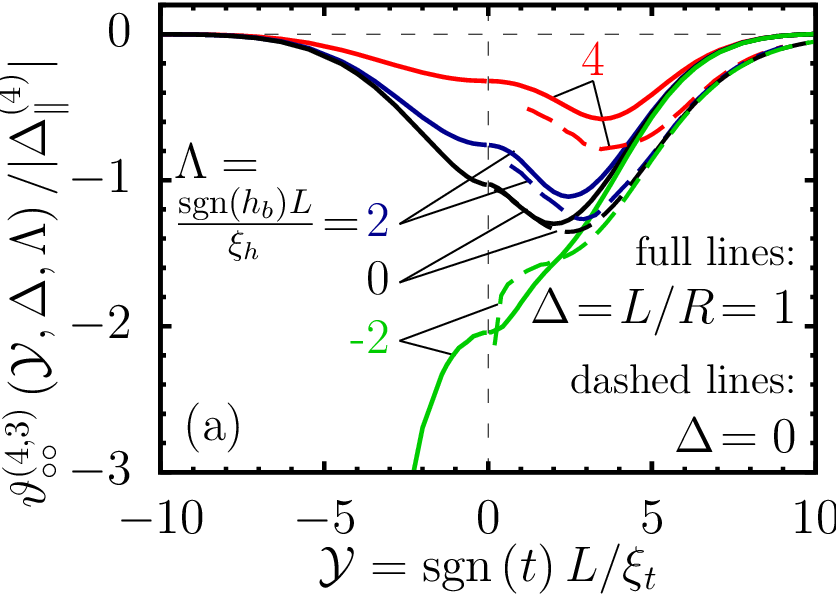}
    \includegraphics[width=0.47\textwidth]{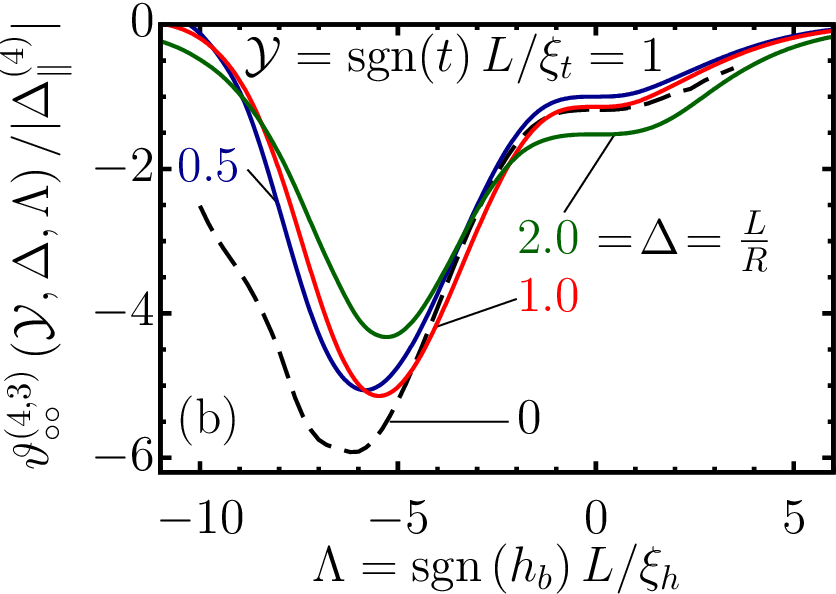}
    \caption{
	Effect of the bulk field \of{\Lambda\neq0} on the scaling function 
	$\sff[\of{\dSpace=4,\dSp=3}]_{\twoSph}$ 
	of the critical Casimir force \erefN{}{eq:ccf_spsp_scaling}{} as obtained 
	from the mean-field theory for the Landau model.
	(a) Normalized $\sff[\of{4,3}]_{\twoSph}$ shown as a function of $\y=\sgn\of{t} L/\xit$ 
	for $\Delta =  L/R=1$ (full lines) 
	and within the Derjaguin approximation ($\Delta = 0$, dashed lines)
	for  four values of $\Lambda=\sgn\of{\hb} L/\xih$.
	(b) Normalized $\sff[\of{4,3}]_{\twoSph}$ shown as a function of 
	$\Lambda$  for $\y=1$ and for four values of $\Delta$.
        The curves are normalized with the critical Casimir amplitude 
        $\Delta_{\film}\lsup{4}$ for the film.}
    \label{fig:H}
\end{figure}

Results for nonzero bulk fields $h_b$ are shown in Fig.~\ref{fig:H}.
For fixed sphere radii $R$ and fixed surface-to-surface distance $L$, 
the curves in Fig.~\ref{fig:H}(a) for fixed $\Lambda$ correspond to varying the 
temperature along the thermodynamic paths of iso-fields $h_b = \mathrm{const}$. 
For fixed $L$, the curves in Fig.~\ref{fig:H}(b) compare the scaling function of the CCF  
as function of \hb{} along the supercritical  isotherm $T_c < T=\mathrm{const}$
for various sphere sizes.

For $\hb>0$ the variation of $\sff[\of{4,3}]_{\twoSph}$ with 
$\y$ resembles the features observed  for vanishing $\hb$ in the case of 
the sphere-sphere or film geometry, i.e., $\sff[\of{4,3}]_{\twoSph}$
exhibits a minimum located above $T_c$ ($\y > 0$) 
[compare \fref{fig:H}(a) with \fref[s]{fig:FvsT_H0}(a) and \ref{fig:sff3d}]. Upon increasing the  
bulk field, the magnitude of the scaling function decreases and the position of the minimum 
shifts towards larger $\y$. This is in line with the behavior 
for the  film geometry (\fref{fig:sff4d_polar}).

The behavior of the scaling function for negative bulk fields is different.
For positive \y, there is still a residual minimum of the scaling function 
located very close to $\y=0$, which disappears upon decreasing \hb{}.
This is already the case  for $\Lambda=-2$ in \fref{fig:H}(a). 
This  disappearance  is in line with the results for film geometry. 
For negative $\y$, at a certain value $\Lambda<0$, in films capillary condensation occurs 
whereas between spherical colloids a bridging transition takes place 
\cite{Archer-et:2005,Okamoto-et:2013,Bauer-et:2000}. Near these phase transitions, 
the effective force acting between the confining 
surfaces is attractive and becomes extremely strong; the depth of the corresponding
effective interaction potentials can reach a few hundred $\kb\T$. 
This concomitant enormous increase of the strength of the force is also reflected 
in the universal scaling function 
[see the green line $\Lambda=-2$ in \fref{fig:H}(a) for $\y<0$]. 
(For the film geometry this issue has 
been discussed in detail in \rcite{Schlesener-et:2003}; in particular, 
Fig.~11 in \rcite{Schlesener-et:2003} exhibits a cusp in the scaling function 
in the vicinity of the capillary condensation; similarly, upon decreasing $\y$, 
called $\Theta_{-}$ in \rcite{Schlesener-et:2003}, to negative values the 
magnitude of the scaling function increases strongly.)

It is also interesting to note a non-monotonic dependence of the scaling function 
$\sff[\of{4,3}]_{\twoSph}$ on $\Delta = L/R$ [\fref{fig:H}(b)]. 
For positive bulk fields, $\abs{\sff[\of{4,3}]_{\twoSph}}$ 
is stronger for larger $\Delta$. This is different, however, for negative bulk fields, 
for which $\abs{\sff[\of{4,3}]_{\twoSph}}$ is stronger for smaller $\Delta$.
Such an increase of \abs{\sff[\of{4,3}]_{\twoSph}} upon decreasing $\Delta$
holds also for zero bulk field [see \fref{fig:FvsT_H0}(b) for $\Delta \gtrsim 1/2$]. 

Finally, for larger values of $\Delta = L/R$ the deficiencies of the Derjaguin 
approximation are clearly visible in Figs.~\ref{fig:FvsT_H0} and \ref{fig:H}.

\section{Comparison with experimental data \label{sec:exp}}

\subsection{Effective interaction potentials \label{sec:exp_pot}}

In \rcite{Dang-et:2013}, the pair distribution function $g\of{r}$ of 
poly-n-isopropyl-acrylamide microgel (PNIPAM) colloidal particles immersed in a 
near-critical 3-methyl-pyridine (3MP)/heavy water mixture has been 
determined experimentally for various deviations $\Delta\T=\tcb-\T$ from the lower 
critical temperature $\tcb\approx39^{\circ}\textrm{C}$ (of the miscibility gap 
of the \emph{bulk} 3MP/heavy water mixture without colloidal particles). 
Here we analyze the experimental data for the 3MP mass fraction $\mf=0.28$ 
which is close to the critical value (see below).

\begin{figure}[!h]
    \includegraphics{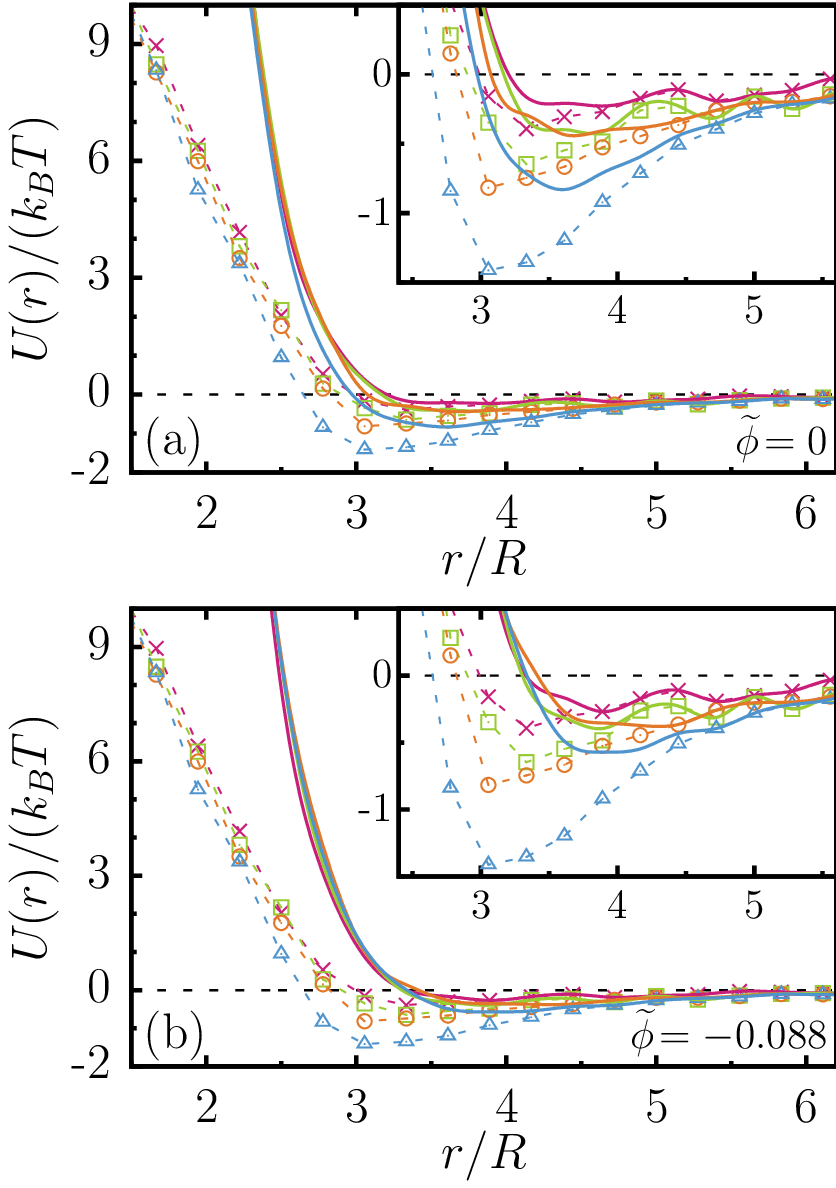}
	\caption{Effective interaction potential \pot[exp] as determined experimentally 
	in \rcite{Dang-et:2013} (symbols, dashed lines as a guide to the eye) 
	and its background contribution \pot[bck] (full lines).
	The experimental system consists of colloidal particles of radius 
	$R \approx 250 \;\mathrm{nm}$ immersed in a near critical binary liquid mixture. 
	The effective potential was determined for various deviations $\Delta\T=\tcb-\T$
	from the lower critical point \tcb, at 
	$\Delta\T/\textrm{K}=0.6$ ($\boldsymbol{\times}$, magenta),  
	$0.5$ ($\boldsymbol{\Box{}}$, green), $0.4$ ($\boldsymbol{\circ}$, orange), 
	and $0.3$ ($\boldsymbol{\bigtriangleup}$, blue).
	Upon approaching \tcb{} the minimum of the potential \pot{} deepens due   
	to the attractive Casimir interaction. The `background' part of the potential 
	is obtained by subtracting the {\it{}c}ritical  Casimir potential 
	\pot[c] \erefN[s]{see }{eq:exp_potbck}{and \eqref{eq:exp_potc}}. 
	If \pot[bck] was temperature independent the various full lines would collapse.  
	In (a) the binary liquid mixture used in the experiments (mass fraction $\mf=0.28$)
	is assumed to be at its critical composition $\mf=\mf_c=0.28$,\ 
	whereas (b) corresponds to a slightly off-critical composition 
	$\widetilde\Op=\of{\mf_c-\mf}/\OpBt=-0.088$. In (b) one observes a better collapse of the 
	full lines than in (a). 
	The critical Casimir potential \pot[c] depends on \Sig, which is directly related 
	to $\widetilde{\Op{}}$ via the equation of state
	$\Sig=\eosSig\of{\abs{t}^{\beta}\widetilde{\Op}}$
	(see the main text). 
	The curves correspond to the value  
	$\xit[+]\lsup{0}=1.5 \;\textrm{nm}$ \erefN{}{eq:crit_corrlength_xit}{}. 
	The colloidal particles are soft, so that $\pot[exp]\of{r<2R}>0$ and very 
	large but not infinite.}
        \label{fig:expcurves}
\end{figure}

We assume that the solvent-mediated interaction between the 
PNIPAM colloids for center-to-center distances $r$ is the sum 
of a {\it{}b}a{\it{}ck}ground contribution \pot[bck] and the {\it{}c}ritical 
Casimir potential \pot[c]. This assumption is valid for small salt concentrations
\cite{pousaneh:jpcm:11, *Bier-et:2011, *Pousaneh-et:2012} which is the case 
for the samples studied in \rcite{Dang-et:2013}.
Accordingly, one has 
\begin{equation}
\label{eq:exp_potbck}
  \pot[bck]\of{r}=\pot[exp]\of{r;\Delta\T}-\pot[c]\of{r;\Delta\T}.
\end{equation}
Within the studied temperature range $\Delta\T < 1\mathrm{K}$
this `background' contribution is expected to depend only weakly on temperature
and hence we consider it to be temperature independent.
We use the potential of mean-force in order to extract the {\it{}exp}erimentally
determined interaction potential $\pot[exp]\of{r}=-\kb\T{}\ln\fd{g\of{r}}$.
This relation is reliable for small solute densities, as they have been used 
in the experiments. Therefore only small deviations are expected to occur 
by using more accurate expressions for the potential, such as the 
hypernetted chain or the Percus-Yevick closures. 

Since  the numerical calculation of the critical Casimir
potential in the bona fide sphere-sphere 
geometry for all parameters which are needed for comparison with experiment 
is too demanding, here we resort to the Derjaguin approximation. 
Within this approximation  
the {\it{}c}ritical Casimir potential \pot[c] between two colloids 
of radius $R$  \erefN[s]{}{eq:ccf_spsp_scaling}{
and \eqref{eq:ccf_derj}}
is \cite{Derjaguin:1934,Hanke-et:1998, Troendle-et:2009, *Troendle-et:2010}
\begin{equation}
\label{eq:exp_potc}
    \pot[c]\of{r;\Delta\T,\mf} 
	=  \pi \kb\T\frac{R}{r-2R} \int_1^{\infty} \mathrm{d}x
	    \of{x^{-2}-x^{-3}}\hat{\vartheta}^{\of{\dSpace=3}}_{\film}\of{x\y,\Sig}, 
\end{equation}
where $\y=\sgn\of{t}\of{r-2R}/\xit$ and $\Sig=\sgn\of{t\hb}\xit/\xih$. 
The dependence of \pot[c] on temperature and on the mass fraction of the solvent 
is captured by the bulk correlation lengths \xit{} and \xih{} of the solvent,
respectively \erefN{}{eq:crit_corrlength}{}. In order to calculate the scaling 
function $\hat{\vartheta}^{\of{\dSpace=3}}_{\film}$ of the critical Casimir force 
between two planar walls we use the local functional approach (see \sref{ssc:th_locfct}).

For the amplitude of the thermal bulk correlation length we take
$\xit[+]\lsup{0} = 1.5 \;\mathrm{nm}$, which we extracted from the experimental data 
presented in \rcite{Sorensen-et:1985}. However, in the literature there are no well established data 
for the critical mass fraction $\mf_c$ of the 3MP/heavy water binary liquid mixtures. 
In \rcite{Cox:1952}, the value $\mf_c=0.28$ is  quoted  while the scaling analysis 
of Fig.~1 in \rcite{Cox:1952} suggests the value $\mf_c \approx 0.29$. The inaccuracy 
of the value for $\mf_c$ enters into the reduced  order parameter
$\widetilde{\Op}=\of{\mf_c-\mf}/\OpBt$; 
\OpBt{} is the non-universal amplitude of the bulk {\it{}c}oe{\it{}x}istence curve 
$\mf_{cx}\of{t=\Delta\T/\tcb<0}=\mf_c\pm\OpBt\abs{t}^{\beta}$. 
Thus,  via the equation of state one obtains $\Sig=\eosSig\of{\abs{t}^{\beta}\widetilde{\Op}}$ (see  
 Eq.~(A4) in the first part of Ref.~\cite{Mohry-et:2012a}) so that 
the critical Casimir potential \pot[c] \erefN{}{eq:exp_potc}{}  depends sensitively on 
  the value of $\mf_c$. 
The function $\eosSig$ is determined by using the equation of state 
within the linear parametric model~\cite{Fisher:1971}.
Note, that as long as we consider the reduced order parameter $\widetilde{\Op{}}$ we do not 
have to know the non-universal amplitude \OpBt{} (or $\xih\lsup{0}$ which is 
related to \OpBt{} via universal amplitude ratios.)

Figure~\ref{fig:expcurves}(a) shows the experimentally determined potentials and the 
extracted background contributions \pot[bck] for the critical composition 
being $\mf_c = 0.28=\mf$, as stated in \rcite{Dang-et:2013}.
In view of the uncertainty in the value of $\mf_c$, we used $\widetilde{\Op}$ as a 
variational parameter for achieving the weakest variation of the `background' 
potential \pot[bck] with temperature.
For example, for $\widetilde{\Op} = -0.088$ the variation of \pot[bck] as 
function of \T{} is smaller than $0.5 k_BT$ and thus comparable with the 
experimentally induced inaccuracy [see  \fref{fig:expcurves}(b)].
For all tested values of $\widetilde{\Op}$, that obtained \pot[bck], which corresponds to 
$\Delta\T/\mathrm{K}=0.2$, deviates the most from the other three curves. 
These deviations might be attributed to the invalidity of the 
Derjaguin approximation (compare \sref{ssc:ccf_spheres}) or  to  the overestimation 
of the CCFs within the local functional approach (compare \fref{fig:sff3d}).
Adopting the value $\OpBt\simeq 0.5$ 
(which can be inferred from the experimental data in \rcite{Cox:1952})
$\widetilde\Op = \of{\mf_c-\mf}/\OpBt= -0.088$ corresponds to a 
critical mass fraction $\mf_c\simeq 0.236$. 
This value of $\mf_c$ differs significantly from the value given in \rcite{Dang-et:2013}. 
We conclude, that either the solvent used in these experiments is 
indeed at the critical composition, but 
\pot[c] does not capture the whole temperature dependence of \pot[exp] [case (a)], or 
\pot[c] does capture the whole temperature dependence of \pot[exp], but 
$\mf=0.28$ is not the critical composition [case (b)]. 
Moreover, also other physical effects, such as a coupling of the 
critical fluctuations  to  electrostatic interactions   
or the structural properties of the soft
microgel particles,
which we have not included in our analysis,
might be of importance for the considered system.

\subsection{Segregation phase diagram \label{sec:exp_phasediagram}}

The experiments of \rcite{Dang-et:2013} indicate that, upon approaching 
the critical point of the solvent,
a colloidal suspension segregates into two phases: poor (\pO) and rich ($\pT$) in colloids. 
Reference \cite{Dang-et:2013} also provides the experimental data for the colloidal packing 
fractions ($\eta_{cx}\lsup{\pO,\pT}$) in the \emph{c}oe\emph{x}isting phases $\pO$ and $\pT$. 
In order to calculate $\eta_{cx}\lsup{\pO,\pT}$, we use the so-called `effective approach,' 
whithin which one considers a one-component system of colloidal particles interacting with 
each other through an effective, solvent-mediated pair potential \pot. Thus this approach 
ignores that the solvent itself may `participate' in the phase separation of the colloidal suspension.
This approximation allows us, however, to make full use of the 
known results of standard liquid state theory
(for more details and concerning the limitations of this approach see 
\rcite[s]{Louis:2002, Mohry-et:2012a}). 

Within the \emph{r}andom-\emph{p}hase \emph{a}pproximation,
the free energy \Fe{} of the effective one-component system 
is given by \cite{Hansen-et:1976,Mohry-et:2012a} 
\begin{equation}
\label{eq:exp_rpaFe}
  \frac{\pi\sigma^3}{6\vol}\Fe[RPA]  
  	= \kb\T \fe[hs]+\frac12 {\eta_{\sigma}}^2 \widetilde{\pot}_{a,0},
\end{equation}
where \vol{} is the volume of the system. For the \emph{h}ard-\emph{s}phere reference 
free energy \fe[hs] we adopt the Percus-Yevick expression 
\begin{equation}
\label{eq:exp_hsEff}
 \fe[hs] /\eta_{\sigma} = 
      {\ln\fd{\frac{\pi}{6}\of{\sigma/\lambda}^3} +\ln\fd{\frac{\eta_{\sigma}}{1-\eta_{\sigma}}}
       -\frac{2-10\eta_{\sigma}+5\eta_{\sigma}^2}{2\of{1-\eta_{\sigma}}^2}},
\end{equation}
where $\eta_{\sigma}=\of{\frac{\sigma}{2R}}^3\eta=\frac{\pi}{6} \sigma^3\varrho$ with 
$\eta$ being the packing fraction of the colloids, $\varrho$ their number density, and 
$\lambda$ is the thermal wavelength. We use for the effective hard-sphere diameter   
$\sigma=\int_0^{r_0}\set{1-\exp\fd{-\pot/(\kb\T)}}\mathrm{d}r $, with $\pot\of{r=r_0}=0$.  
One can  adopt also other  definitions of  $\sigma$ 
(for a  discussion see Refs.~\cite{Andersen-et:1971,NoteHsdiameter}).
Using the present definition renders a slightly better agreement with the
experimental data than using the  one given in Ref.~\cite{NoteHsdiameter}.

In \eref{eq:exp_rpaFe}, one has  $\widetilde{\pot}_{a,0}=\frac6{\pi\sigma^3}\ft{\pot}_a\of{q=0}$, 
where $\ft{\pot}_a\of{q=\abs{\vc{q}}} = \int \exp\of{-i\vc{q}\vc{r}}\pot[a]\of{r}\mathrm{d}^3{r}$ 
is the Fourier transform of the \emph{a}ttractive part (\pot[a]) of the interaction potential,
\begin{equation}
\label{eq:exp_ua}
  \pot[a]\of{r}=
    \begin{cases}
        \pot\of{r=r_{min}} & \text{ for } 0 \leq r< r_{min} \\
        \pot\of{r}         & \text{ for } r \geq r_{min}    ,
    \end{cases}
\end{equation}
where $\pot\of{r}$ attains its minimum at $r_{min}$. 

\begin{figure}
        \includegraphics{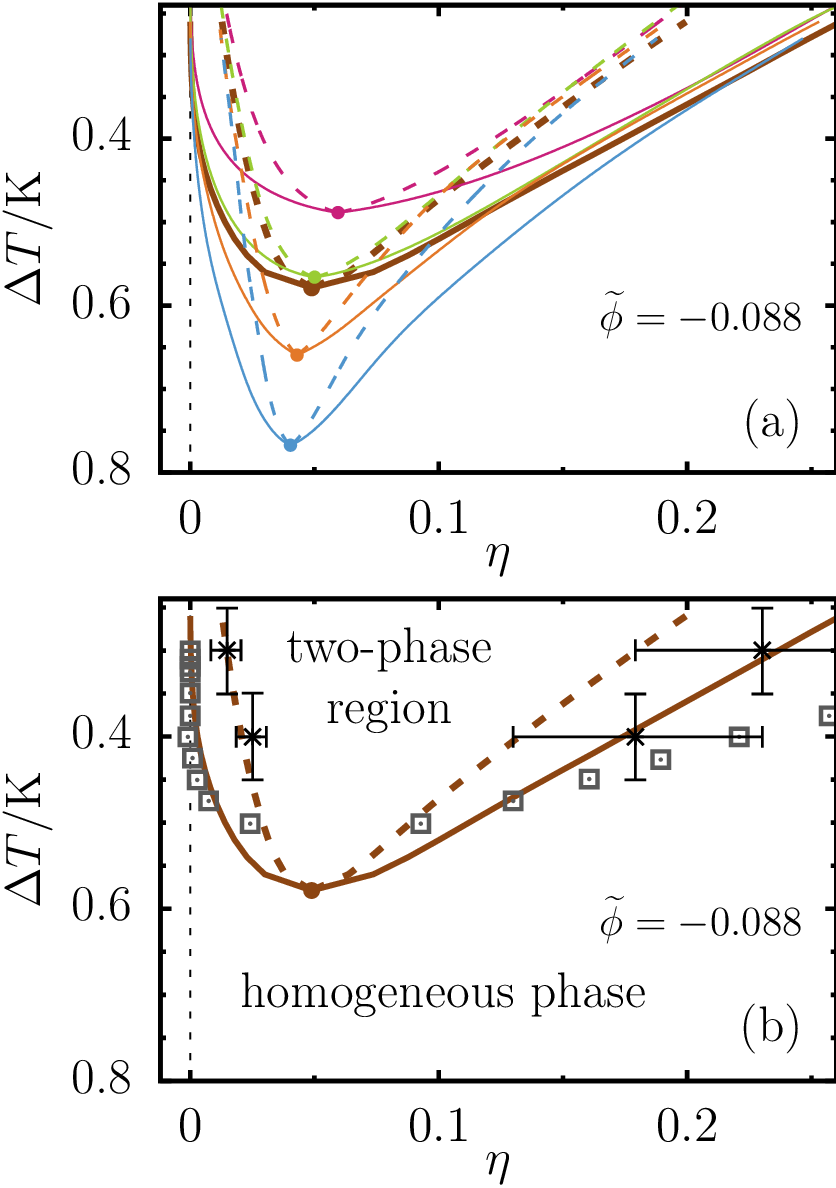}
        \caption{
	Segregation phase diagram from theory (RPA), experiment, and simulations (MC). 
	(a)~The phase diagram obtained within RPA using the four available background 
	potentials \pot[bck] from \fref{fig:expcurves}, and their average. The 
	critical Casimir potential is calculated within the Derjaguin approximation 
	using the local functional approach (see Sec.~\ref{sec:exp_pot}) 
	for a reduced solvent order parameter $\widetilde\Op = -0.088$. 
	The background contributions \pot[bck] have been 
	extracted from the experimentally determined effective potentials 
	\erefN{compare }{eq:exp_potbck}{} at $\Delta\T/\textrm{K}=0.6$ (magenta), 
	$0.5$ (green), $0.4$ (orange), and $0.3$ (blue). The thick dark red curve corresponds 
	to the average of the four potentials \pot[bck]. The solid lines show the phase 
	boundaries in terms of the packing fraction $\eta$ of the colloids, the dashed 
	lines correspond to the spinodals, and dots represent critical points. 
	(b)~Comparison of the theoretical predictions for the phase boundaries (based
	on the average \pot[bck]) with Monte Carlo simulations ($\boxdot$) and experiments 
	($\boldsymbol{\times}$, with error bars) of Ref.~\cite{Dang-et:2013}. 
        On the temperature axis $\Delta{}T=\tcb-\T$ 
	increases from top to bottom in order to mimic a lower critical point \tcb{} 
	(of the solvent) as observed experimentally.}
        \label{fig:exp_phasediagram1}
\end{figure}

In order to calculate the phase diagram of the effective one-component 
system within the RPA approximation, we use the pair potential 
$\pot\of{r}=\pot[bck]\of{r}+\pot[c]\of{r}$, where \pot[c] is given by 
\eref{eq:exp_potc}, and where the background contribution \pot[bck] is 
extracted from the experimental data of \rcite{Dang-et:2013}.  
As discussed in \sref{sec:exp_pot}, there is some inaccuracy in determining 
the background potential \pot[bck].
Following \rcite{Dang-et:2013} and assuming $\widetilde\Op = 0$, we have to
consider four different \pot[bck]. The resulting corresponding segregation phase diagrams   
differ from each other qualitatively.   
Interestingly, the attractive part of the background potentials 
$\pot[bck]\of{r; \Delta\T,\widetilde\Op}$ corresponding to $\Delta\T/\mathrm{K}=0.4$ 
and $0.2$ [see \fref{fig:expcurves}(a)] is so strong, that for these potentials 
alone (i.e., for $\pot=\pot[bck]$ without \pot[c]) the RPA free energy predicts 
already a phase segregation. 
For the background potential $\pot[bck]\of{r; \Delta\T,\widetilde\Op}$ 
corresponding to $\Delta\T=0.6\mathrm{K}$ and $\widetilde\Op=0$, 
the presence of \pot[c] is necessary for the occurrence of phase segregation within RPA.  
However, the resulting relative value of the critical temperature 
$(\Delta\T)_{c,eff}\simeq 0.39\textrm{K}$ 
is much smaller than the experimentally observed one.
On the other hand, for $\widetilde\Op = -0.088$, which renders the best expression for \pot[bck] out of 
the experimental data of \rcite{Dang-et:2013} (see Fig.~\ref{fig:expcurves}), 
the resulting RPA phase segregation diagrams are consistent with each other. 
This is visible in \fref{fig:exp_phasediagram1}(a),  
where we compare the {\it{}c}oe{\it{}x}istence 
curves $\eta_{cx}\of{\T}$ resulting from the four potentials \pot[bck] of 
Fig.~\ref{fig:expcurves}(b), as well as from \pot[bck] obtained by averaging 
these four potentials \pot[bck].
Although these five background potentials look 
very similar, they nonetheless lead to coexistence curves the critical temperatures 
of which differ noticeably [see \fref{fig:exp_phasediagram1}(a)]. 
However, away from their critical point, the  various  coexistence 
curves merge; see the region $\Delta\T<0.4\mathrm{K}$ in \fref{fig:exp_phasediagram1}(a).
This indicates that for small $\Delta T$ 
the critical Casimir potential dominates the background potential, so that the 
details of the latter (and thus its inaccuracy) become less important.

Figure~\ref{fig:exp_phasediagram1}(b) compares the RPA predictions for the 
segregation phase diagram with the experimental data and with the Monte Carlo 
simulation data provided by Ref.~\cite{Dang-et:2013}. 
The pair potentials used in these MC simulations  are the   sum of an attractive 
and  a  repulsive exponential function and thus  they
 differ from the ones used here.
At high colloidal densities, 
the RPA is in surprisingly good agreement with the experimental data. On the 
other hand, at low densities the RPA agrees well with the Monte Carlo simulations, 
while there both underestimate the experimental values which, in turn, agree well 
with the RPA-spinodal (an observation also observed for $\widetilde\Op=0$). 
While this latter `agreement' might be accidental, it 
nevertheless raises the question whether the experimental system has actually been 
fully equilibrated at the time of the measurements.

\section{Summary \label{sec:summary}}

Critical Casimir forces act between surfaces confining a near-critical medium. For instance, 
colloidal particles suspended in a binary liquid mixture act as cavities in
this solvent. Thus near its critical point of demixing the suspended colloids interact via 
an effective, solvent-mediated force, the so-called critical Casimir force
(CCF). We have analyzed the dependence of the CCFs on the {\it{}b}ulk ordering 
field (\hb{}) conjugate to the order parameter of the solvent. For a binary liquid
mixture, \hb{} is proportional to the deviation of the difference of the chemical potentials of the
two species from its critical value. In the presence of \hb, we have 
used mean-field theory to calculate the CCFs between parallel
plates and between two spherical colloids, as well as the local functional  approach
of Fisher and de Gennes for parallel plates. We have shown that the CCF is
asymmetric around the consolute point of the solvent, and that it is stronger
for compositions slightly poor in that species of the mixture which
preferentially adsorbs at the surfaces of the colloids (see
Figs.~\ref{fig:sff4d_polar}, \ref{fig:sff3d}(a), \ref{fig:sff3d}(b), and \ref{fig:H}).

For two three-dimensional spheres posing
as hypercylinders $(H_{3,4})$ in spatial dimension $D=4$  
we observe a non-monotonic dependence of the scaling function of the  CCF 
on the scaling variable $\Delta = L/R$, where $L$ is the surface-to-surface
distance and $R$ is the
radius of monodisperse colloids [see Fig.~\ref{fig:FvsT_H0}(b) as well as 
Fig.~\ref{fig:H}]. 
Unlike four-dimensional spheres ($H_{4,4}$) in $D=4$, the scaling functions
for $H_{3,4}$ exhibit a maximum at $\Delta \approx 1/2$ before decreasing
upon increasing $\Delta$ [see Fig.~\ref{fig:FvsT_H0}(b)]. This different behavior 
may be attributed to the extra macroscopic extension of the hypercylinders $H_{3,4}$. 
This raises the question whether $H_{3,4}$ or $H_{4,4}$ is the better
mean-field approximation
for the physically relevant case of three-dimensional spheres $H_{3,3}$ in $D=3$. 
Due to this uncertainty and also in view of the  limited
reliability of the Derjaguin approximation (see Figs.~\ref{fig:FvsT_H0} and
\ref{fig:H}) more accurate theoretical approaches are highly desirable.
Because the local functional approach is computational less demanding than Monte Carlo  
simulations and it is reliable for $\hb=0$, it would be very useful 
to improve this approach for $\hb\neq0$ and to generalize it to 
more complex geometries, in particular to spherical objects. 

In addition, due to numerical difficulties the behavior of the scaling
function of the CCF for $\Delta \to \infty$ remains as an open issue. 
Since one faces similar
numerical difficulties for $\Delta \to 0$, we conclude that within  mean-field
theory the numerical solution finds its useful place in between small and large
colloid separations. The small separations are captured well by the
Derjaguin approximation. For $H_{d,D}$ spheres with $\dSp>\beta\dSpace/\nu$,
the large separations can be investigated by the so-called small radius expansion. 
However, the case $H_{3,4}$ represents a `marginal' perturbation for which the 
small radius expansion is not valid~\cite{Hanke-et:1999}. Therefore it would be 
interesting to study the asymptotic behavior of the scaling function of the CCF 
for large colloid separations by other means.

We have compared our theoretical results for the critical Casimir potential 
\erefN{within the Derjaguin approximation and the local functional approach,
see }{eq:exp_potc}{} with experimental data taken from \rcite{Dang-et:2013}  
(see \fref{fig:expcurves}). 
Concerning the potentials we find a fair agreement, however 
their detailed behavior calls for 
further, more elaborate experimental and theoretical investigations. 

As a  consequence of the emergence of CCFs, a colloidal suspension 
thermodynamically close to the critical point of its solvent 
undergoes phase separation into a phase dense in colloids and a phase dilute in colloids. Using
the random phase approximation for an effective one-component system, 
we have calculated the phase 
diagram for this segregation in terms of the colloidal packing fraction and 
of the deviation of temperature from that of the critical point of the solvent. 
Surprisingly, despite resorting to these approximations, the
calculated phase diagram agrees fairly well with the corresponding
experimental and Monte Carlo data (Fig.~\ref{fig:exp_phasediagram1}). 
Both the RPA calculations and the Monte Carlo simulations 
are based on the so-called effective approach and compare 
similarly well with the experimental data. 
However, in order to achieve an even better agreement with the experimental data, it  
is likely that models have to be considered 
which take into account the truely 
ternary character of the colloidal suspension.

\begin{acknowledgments}
We thank M. T. Dang, V. D. Nguyen, and P. Schall for interesting
discussions about their experiments and for providing us their data. 
\end{acknowledgments}
%
%
%
%
%
%
%

\begin{thebibliography}{70}
\bibitem [{Not({\natexlab{b}})}]{NoteHsdiameter}%
  \BibitemOpen
  \href@noop {} {}\bibinfo {note} {We have also considered
  the definition
  $\sigma=\int_0^{r_{min}}\set{1-\exp\fd{-\pot[r]/(\kb\T)}}\mathrm{d}r$, where
  $\pot[r](r)=\pot(r)-\pot(r_{min})$ with $\pot\of{r}$ attaining its minimum at
  $r_{min}$.}\BibitemShut {Stop}%
\end{thebibliography}
\end{document}